\documentclass[journal]{IEEEtran}
\usepackage[english]{babel}
\usepackage{amsmath}
\usepackage{amsthm}
\usepackage{amsfonts}
\usepackage{cite}
\usepackage{hyperref}
\newtheorem{thm}{Theorem}

\newtheorem*{thm*}{Theorem}
\newtheorem{cor}{Corollary}

\newtheorem{lem}{Lemma}

\newtheorem{prop}{Proposition}

\theoremstyle{definition}
\newtheorem{defn}{Definition}

\newtheorem{ex}{Example}

\theoremstyle{remark}
\newtheorem{rem}{Remark}

\begin{document}
\title{The Quantum Wasserstein Distance of Order 1}
\author{Giacomo De Palma, Milad Marvian, Dario Trevisan, and Seth Lloyd%
\thanks{Manuscript received December 2, 2020; accepted April 15, 2021. The work
of Giacomo De Palma and Seth Lloyd was supported by the U.S. Air Force
Office of Scientific Research, the U.S. Army Research Office through the
Blue Sky Program, DOE, and IARPA. The work of Milad Marvian was
supported in part by the U.S. Air Force Office of Scientific Research, the
U.S. Army Research Office through the Blue Sky Program, DOE, and IARPA,
and in part by the NSF under Grant CCF-1954960. The work of Dario
Trevisan was supported in part by the University of Pisa under Project PRA
2018-49 and in part by the INdAM through the GNAMPA Project 2020
Problemi di ottimizzazione con vincoli via trasporto ottimo e incertezza. This
article was presented in part at the Entropy Inequalities, Quantum Information
and Quantum Physics workshop, the 24th Annual Conference on Quantum
Information Processing (QIP 2021), the Beyond IID in Information Theory 8
workshop, and the Young Italian Quantum Information Science Conference
(IQIS) 2020.}
\thanks{Giacomo De Palma was with the Research Laboratory of Electronics and the
Department of Mechanical Engineering, Massachusetts Institute of Technology,
Cambridge, MA 02139 USA. He is now with the Scuola Normale
Superiore, 56126 Pisa, Italy (e-mail: giacomo.depalma@sns.it).}
\thanks{Milad Marvian was with the Research Laboratory of Electronics and the Department
of Mechanical Engineering, Massachusetts Institute of Technology,
Cambridge, MA 02139 USA. He is now with the Center for Quantum
Information and Control (CQuIC), Department of Electrical and Computer
Engineering, University of New Mexico, Albuquerque, NM 87131 USA.}
\thanks{Dario Trevisan is with the Mathematics Department, Università degli Studi
di Pisa, 56127 Pisa, Italy.}
\thanks{Seth Lloyd is with the Research Laboratory of Electronics and the Department of
Mechanical Engineering, Massachusetts Institute of Technology, Cambridge,
MA 02139 USA.}
\thanks{Communicated by S. Beigi, Associate Editor for Quantum Information
Theory.}
\thanks{Digital Object Identifier 10.1109/TIT.2021.3076442}
}

\markboth{IEEE Transactions on Information Theory}%
{De Palma \MakeLowercase{\textit{et al.}}: The quantum Wasserstein distance of order 1}

\maketitle

\begin{abstract}
We propose a generalization of the Wasserstein distance of order 1 to the quantum states of $n$ qudits.
The proposal recovers the Hamming distance for the vectors of the canonical basis, and more generally the classical Wasserstein distance for quantum states diagonal in the canonical basis.
The proposed distance is invariant with respect to permutations of the qudits and unitary operations acting on one qudit and is additive with respect to the tensor product.
Our main result is a continuity bound for the von Neumann entropy with respect to the proposed distance, which significantly strengthens the best continuity bound with respect to the trace distance.
We also propose a generalization of the Lipschitz constant to quantum observables.
The notion of quantum Lipschitz constant allows us to compute the proposed distance with a semidefinite program.
We prove a quantum version of Marton's transportation inequality and a quantum Gaussian concentration inequality for the spectrum of quantum Lipschitz observables.
Moreover, we derive bounds on the contraction coefficients of shallow quantum circuits and of the tensor product of one-qudit quantum channels with respect to the proposed distance.
We discuss other possible applications in quantum machine learning, quantum Shannon theory, and quantum many-body systems.
\end{abstract}

\begin{IEEEkeywords}
Quantum optimal mass transport, Wasserstein distance, Hamming distance, qudits, von Neumann entropy, Lipschitz constant, concentration inequalities.
\end{IEEEkeywords}

\section{Introduction}

\subsection{Motivations}
\IEEEPARstart{T}{he} most prominent distinguishability measures between quantum states are the trace distance, the quantum fidelity and the quantum relative entropy, and they all have in common the property of being unitarily invariant \cite{nielsen2010quantum,wilde2017quantum,holevo2019quantum}.
A fundamental consequence of this property is that the distance between any couple of quantum states with orthogonal supports is always maximal.
However, this property is not always desirable.
For certain applications, it is natural to use a distance with respect to which the state $|0\rangle^{\otimes n}$ is much closer to $|1\rangle\otimes|0\rangle^{\otimes\left(n-1\right)}$ than to $|1\rangle^{\otimes n}$.
Some desirable properties can be recovering the Hamming distance for vectors of the canonical basis, and more generally robustness against local perturbations on the input states.
Such a distance may, for example, provide better continuity bounds for the von Neumann entropy since the von Neumann entropy is also robust against local perturbations. In particular, any operation on one qubit can change the entropy of a state by at most $\ln4$, which does not depend on the number of qubits.
Therefore, the entropy of an $n$-qubit state with initial entropy $O(n)$ remains $O(n)$ after such an operation. However, this continuity property cannot be captured by any unitarily invariant distinguishability measure, since a one-qubit operation can bring the initial state into an orthogonal state, resulting in a maximum possible change in the unitarily invariant measure.

\subsection{The classical Wasserstein distances}
In the setting of classical probability distributions on a metric space, the distances originating from the theory of optimal mass transport have emerged as prominent distances with the properties above.
Their exploration has led to the creation of an extremely fruitful field in mathematical analysis, with applications ranging from differential geometry and partial differential equations to machine learning \cite{villani2008optimal, ambrosio2008gradient, peyre2019computational}.

Given a finite set $\mathcal{X}$, any distance $D$ on $\mathcal{X}$ induces a transport distance on the set of the probability distributions on $\mathcal{X}$, where the distance between the probability distributions $p$ and $q$ is the minimum of the mean distance over joint probability distributions on $\mathcal{X}^2$ with marginals $p$ and $q$. More precisely, we have the following definitions:

\begin{defn}[Coupling]
A coupling between the probability distributions $p$ and $q$ on $\mathcal{X}$ is a probability distribution $\pi$ on $\mathcal{X}^2$ with marginals $p$ and $q$, \emph{i.e.}, such that
\begin{equation}\label{eq:coupling-constraints}
p(x) = \sum_{y\in\mathcal{X}}\pi(x,y)\,,\quad q(y) = \sum_{x\in\mathcal{X}}\pi(x,y)\,,\quad x,\,y\in\mathcal{X}\,.
\end{equation}
We denote with $\mathcal{C}(p,q)$ the set of the couplings between $p$ and $q$.
\end{defn}

\begin{defn}[Classical $W_\alpha$ distances]
For any $\alpha\ge1$, the $W_\alpha$ distance or Wasserstein distance of order $\alpha$ between the probability distributions $p$ and $q$ on $\mathcal{X}$ is
\begin{equation}\label{eq:alpha-wasserstein}
W_\alpha(p,q) = \left(\min_{\pi\in\mathcal{C}(p,q)}\sum_{x,\,y\in\mathcal{X}}{D(x,y)}^\alpha\,\pi(x,y)\right)^\frac{1}{\alpha}\,.
\end{equation}
\end{defn}

Although many properties of the $W_\alpha$ distances do not depend on the choice of $\alpha$, in recent years the distances $W_1$ and $W_2$ are playing a prominent role. The $W_1$ distance is also called Monge--Kantorovich distance, after the foundational works of Monge and Kantorovich\cite{monge1781memoire, kantorovich1942translocation}. In particular, Kantorovich noticed that the $W_1$ distance is in fact induced by a norm, and introduced the transport problem \eqref{eq:alpha-wasserstein} as a linear programming problem (see \cite{vershik2013long} for a detailed historical account).

In many cases the set $\mathcal{X}$ is already endowed with a distance, \emph{e.g.}, when dealing with subsets of Riemannian manifolds or weighted graphs. However, one can always consider the trivial distance
\begin{equation}
D(x,y) = \left\{
           \begin{array}{l}
             0\qquad x=y \\
             1\qquad x\neq y \\
           \end{array}
         \right.
\,,
\end{equation}
and the induced $W_1$ distance coincides with the total variation distance
\begin{equation}
W_1(p,q) = \frac{1}{2}\left\|p-q\right\|_1\,.
\end{equation}

The Hamming distance provides a natural choice when $\mathcal{X}$ is a set of finite strings over an alphabet:

\begin{defn}[Hamming distance]
For any $k\in\mathbb{N}$, let $[k] = \left\{1,\,\ldots,\,k\right\}.$ The Hamming distance between $x,\,y\in[d]^n$ is the number of different components:
\begin{equation}
h(x,y) = \left|\left\{i\in[n]:x_i\neq y_i\right\}\right|\,.
\end{equation}
\end{defn}
The classical $W_1$ distance with respect to the Hamming distance is called Ornstein's $\bar{d}$ distance and was first considered in \cite{ornstein1973application}, together with its extension to stationary stochastic processes. It has found many applications in ergodic theory and information theory, such as coding theorems for a large class of discrete noisy channels with memory and rate distortion theory \cite{gray2013entropy}.

Finally, there has been a surge of interest towards applications of transportation distances in machine learning in the context of Generative Adversarial Networks (GANs) \cite{arjovsky2017wasserstein}.
GANs \cite{goodfellow2014generative} provide a useful algorithm to learn an unknown probability distribution using a neural network. The learning is performed by training a generator trying to produce samples of the unknown distribution against a discriminator trying to distinguish the true from the generated samples.
The training process is a minimax game that converges to a Nash equilibrium. The choice of the loss functions for the discriminator plays a crucial role to ensure convergence in the training procedure.
Employing the Wasserstein distances as loss function of GANs alleviates the problem of the vanishing gradient in the training, which plagued the original version with the Jensen--Shannon divergence (the symmetrized relative entropy)\cite{arjovsky2017wasserstein}.
Suitable variants of the Wasserstein distances which further improve the efficiency of the training have also been proposed \cite{gulrajani2017improved, genevay2018learning}.

\subsection{Our contribution}\label{sec:contr}
We propose a generalization of the $W_1$ distance to the set of the quantum states of $n$ qudits.
The proposed quantum $W_1$ distance is based on the notion of neighboring states.
We anticipate here an informal definition and refer to \autoref{sec:dist} for the details.

\begin{defn}[Quantum $W_1$ distance, informal]
Two quantum states of $n$ qudits are neighboring if they coincide after a suitable qudit is discarded.
The quantum $W_1$ distance is the maximum distance that is induced by a norm that assigns distance at most one to any couple of neighboring states.
\end{defn}

In \autoref{sec:prop}, we prove several properties of the proposed quantum $W_1$ distance:
\begin{itemize}
\item Its ratio with the trace distance lies between $1$ and $n$ (\autoref{sec:tr}).
\item It is invariant with respect to permutations of the qudits and unitary operations acting on one qudit (\autoref{sec:symm}) and additive with respect to the tensor product (\autoref{sec:tens}).
    Moreover, the $W_1$ distance between two quantum states which coincide after discarding $k$ qudits is at most $2k$ (\autoref{sec:marg}).
In particular, any quantum operation on $k$ qudits can displace the initial quantum state by at most $2k$ in the proposed distance.

\item It recovers the Hamming distance for vectors of the canonical basis, and more generally the classical $W_1$ distance for quantum states diagonal in the canonical basis  (\autoref{sec:class}).
\end{itemize}

In \autoref{sec:dual}, we define a generalization to quantum observables of the Lipschitz constant of real-valued functions on a metric space.
We prove that, as in the classical case, the proposed quantum $W_1$ distance between two quantum states is equal to the maximum difference between the expectation values of the two states with respect to an observable with Lipschitz constant at most one.
This dual formulation provides a recipe to calculate the proposed quantum $W_1$ distance using a semidefinite program.

Our main result is a continuity bound for the von Neumann entropy with respect to the proposed quantum $W_1$ distance (\autoref{sec:ent}).
In the limit $n\to\infty$ this bound implies that, if two quantum states have distance $o(n/\ln n)$, their entropies can differ by at most $o(n)$.
The von Neumann entropy is intimately linked to the entanglement properties of a quantum state, and our bound implies that the entanglement of a quantum state is robust against perturbations with size $o(n/\ln n)$ in the quantum $W_1$ distance.

In \autoref{sec:Marton}, we explore the relation between the quantum $W_1$ distance and the quantum relative entropy. In particular, we prove a quantum generalization of Marton's trasportation inequality, stating that the square root of the relative entropy between a generic quantum state and a product quantum state provides an upper bound to their quantum $W_1$ distance.
In \autoref{sec:conc}, we apply the quantum Marton's inequality to prove an upper bound to the partition function of a quantum Hamiltonian in terms of its quantum Lipschitz constant.
A fundamental consequence of this result is a quantum Gaussian concentration inequality, stating that most of the eigenvalues of a quantum observable lie in a small interval whose size depends on its Lipschitz constant.

In \autoref{sec:ch}, we study the contraction coefficient with respect to the proposed quantum $W_1$ distance of the $n$-th tensor power of a one-qudit quantum channel.
While the contraction coefficient of these quantum channels with respect to the trace distance is trivial in the limit $n\to\infty$, we are able to prove an upper bound to the contraction coefficient for the proposed quantum $W_1$ distance which does not depend on $n$.
Moreover, we prove that the contraction coefficient of a generic $n$-qudit quantum channel with respect to the proposed quantum $W_1$ distance is upper bounded by the size of the light-cones of the qudits.

We conclude in \autoref{sec:appl} by discussing other possible applications of the defined quantum $W_1$ distance in quantum machine learning, quantum information, and quantum many-body systems.

\subsection{Related works}
Several quantum generalizations of the Wasserstein distances have been proposed.
One line of research, initiated by Carlen, Maas, Datta and Rouz\'e \cite{carlen2014analog,carlen2017gradient,carlen2020non,rouze2019concentration,datta2020relating,van2021geometrical,gao2020fisher}, defines a quantum $W_2$ distance built on the definition of a quantum differential structure and on the equivalent dynamical formulation of the $W_2$ distance provided by Benamou and Brenier \cite{benamou2000computational}, which assigns a length to each path of probability distributions that connects the source with the target.
The key property of this proposal is that the resulting quantum distance is induced by a Riemannian metric on the manifold of quantum states, and the quantum generalization of the heat semigroup is the gradient flow of the von Neumann entropy with respect to this metric.
This quantum generalization of the $W_2$ distance has been shown to be intimately linked to both entropy and Fisher information \cite{datta2020relating}, and has led to determine the rate of convergence of the quantum Ornstein-Uhlenbeck semigroup \cite{carlen2017gradient,de2018conditional}.
Exploiting their quantum differential structure, Refs. \cite{rouze2019concentration,carlen2020non,gao2020fisher} also define a quantum generalization of the Lipschitz constant and the $W_1$ distance, and prove that it satisfies a Talagrand inequality, which also implies some concentration inequalities.
Alternative definitions of quantum $W_1$ distances based on a quantum differential structure are proposed in Refs. \cite{chen2017matricial,ryu2018vector,chen2018matrix,chen2018wasserstein}.
Refs. \cite{agredo2013wasserstein,agredo2016exponential,ikeda2020foundation} propose quantum $W_1$ distances based on a distance between the vectors of the canonical basis.

Another line of research by Golse, Mouhot, Paul and Caglioti \cite{golse2016mean,caglioti2021towards,golse2018quantum,golse2017schrodinger,golse2018wave,caglioti2019quantum} arose in the context of the study of the semiclassical limit of quantum mechanics and defines a quantum $W_2$ distance built on a quantum generalization of the couplings.
This distance was the key element to prove that the mean-field limit of quantum mechanics is uniform in the semiclassical limit \cite{golse2016mean}, and has been employed as a cost function to train the quantum counterpart of deep generative adversarial networks \cite{lloyd2018quantum,chakrabarti2019quantum}.
Ref. \cite{de2021quantum} proposes another quantum $W_2$ distance based on quantum couplings, with the property that each quantum coupling is associated to a quantum channel.
The relation between quantum couplings and quantum channels in the framework of von Neumann algebras has been explored in \cite{duvenhage2018balance}.
The problem of defining a quantum $W_1$ distance through quantum couplings has been explored in Ref. \cite{agredo2017quantum}.

The quantum $W_\alpha$ distance between two quantum states can be defined as the classical $W_\alpha$ distance between the probability distributions of the outcomes of an informationally complete measurement performed on the states, which is a measurement whose probability distribution completely determines the state.
This definition has been explored for Gaussian quantum systems with the heterodyne measurement in Refs. \cite{zyczkowski1998monge,zyczkowski2001monge,bengtsson2017geometry}.

Notions of quantum Hamming ball of a subspace have been defined in Refs. \cite{osborne2009quantum,vidick2016simple}, who employ them to prove a Talagrand concentration inequality and a quantum generalization of de Finetti's theorem, respectively.

The Wasserstein distances have also been generalized to other noncommutative settings, such as noncommutative geometry, where they are related to Connes' spectral distances \cite{connes1992metric, cagnache2011spectral, martinetti2018connes}, with applications to  convergence problems of noncommutative spaces \cite{rieffel2004compact, rieffel2004gromov, d2013metric}. Generalizations of the Wasserstein distances have been proposed in free probability, with applications in random matrix theory \cite{biane2001free}.

\section{Notation}
Let $\left\{|1\rangle,\,\ldots,\,|d\rangle\right\}$ be the canonical basis of $\mathbb{C}^d$, and $\mathcal{H}_n = \left(\mathbb{C}^d\right)^{\otimes n}$ be the Hilbert space of $n$ qudits.
We denote by $\mathcal{O}_n$ the set of the self-adjoint linear operators on $\mathcal{H}_n$, by $\mathcal{O}_n^T\subset\mathcal{O}_n$ the subset of the traceless self-adjoint linear operators on $\mathcal{H}_n$, by $\mathcal{O}_n^+\subset\mathcal{O}_n$ the subset of the positive semidefinite linear operators on $\mathcal{H}_n$, by $\mathcal{S}_n\subset\mathcal{O}_n^+$ the set of the quantum states of $\mathcal{H}_n$, and by $\mathcal{P}_n$ the set of the probability distributions on $[d]^n$.
For any $\mathcal{I}\subseteq[n]$, let $\rho_{\mathcal{I}}$ be the marginal of $\rho\in\mathcal{S}_n$ over the qudits in $\mathcal{I}$.
For any $X\in\mathcal{O}_n$, let $\left\|X\right\|_1$ be its trace norm, given by the sum of the absolute values of its eigenvalues.

\section{The quantum \texorpdfstring{$W_1$}{W1} distance}\label{sec:dist}
Our proposal for a quantum Wasserstein distance of order 1 is based on the following notion of neighboring quantum states:

\begin{defn}[Neighboring quantum states]\label{defn:neigh}
We say that $\rho$ and $\sigma\in\mathcal{S}_n$ are neighboring if they coincide after discarding one qudit, \emph{i.e.}, if $\mathrm{Tr}_i\rho = \mathrm{Tr}_i\sigma$ for some $i\in[n]$.
We denote by $\mathcal{N}_n\subset\mathcal{O}_n^T$ the set of the differences between couples of neighboring quantum states:
\begin{align}
&\mathcal{N}_n = \bigcup_{i=1}^n\mathcal{N}_n^{(i)}\,,\nonumber\\
&\mathcal{N}_n^{(i)}=\left\{\rho - \sigma : \rho,\,\sigma\in\mathcal{S}_n,\,\mathrm{Tr}_i\rho = \mathrm{Tr}_i\sigma\right\}\,,\qquad i\in[n]\,,
\end{align}
and with
\begin{align}
\mathcal{B}_n &= \left\{\sum_{i=1}^n p_i\left(\rho^{(i)} - \sigma^{(i)}\right): p_i\ge0,\right.\nonumber\\
& \phantom{=} \quad \left.\sum_{i=1}^np_i=1,\;\rho^{(i)},\,\sigma^{(i)}\in\mathcal{S}_n,\;\mathrm{Tr}_i\rho^{(i)}=\mathrm{Tr}_i\sigma^{(i)}\right\}
\end{align}
the convex hull of $\mathcal{N}_n$.
\end{defn}

\begin{rem}
Other equivalent definitions of neighboring quantum states are possible, see \autoref{app:neigh} for details.
\end{rem}

\begin{prop}
$\mathcal{B}_n$ is a bounded, closed, centrally symmetric (\emph{i.e.}, $-\mathcal{B}_n = \mathcal{B}_n$) and convex subset of $\mathcal{O}_n^T$ with nonempty interior.
\begin{proof}
The only nontrivial property is the nonempty interior, which follows from \autoref{prop:1T}, where we will prove that $\mathcal{B}_n$ is the unit ball of a norm which is upper bounded by $n/2$ times the trace norm.
\end{proof}
\end{prop}

The classical $W_1$ distance is induced by a norm, and the distance between any couple of neighboring probability distributions on $[d]^n$ is at most one (\autoref{lem:nc} of \autoref{app:lemmas}).
Therefore, we look for a distance between quantum states that is induced by a norm that assigns distance at most one to each couple of neighboring quantum states, \emph{i.e.}, $\mathcal{N}_n$ should be contained in the unit ball of the norm.
Since the unit ball of any norm is convex, also $\mathcal{B}_n$ should be contained in the unit ball of the norm.
Any norm is completely determined by its unit ball, and any bounded, closed, centrally symmetric and convex set with nonempty interior is the unit ball of some norm.
Therefore, we define the quantum $W_1$ norm as the unique norm on $\mathcal{O}_n^T$ whose unit ball is $\mathcal{B}_n$:

\begin{defn}[Quantum $W_1$ norm]
We define the quantum $W_1$ norm on $\mathcal{O}_n^T$ as the unique norm with unit ball $\mathcal{B}_n$, \emph{i.e.}, for any $X\in\mathcal{O}_n^T$,
\begin{align}\label{eq:defW1}
\left\|X\right\|_{W_1} &= \min\left(t\ge0:X\in t\,\mathcal{B}_n\right)\nonumber\\
&= \frac{1}{2}\min\left(\sum_{i=1}^n\left\|X^{(i)}\right\|_1: X^{(i)}\in\mathcal{O}_n^T,\,\mathrm{Tr}_i\,X^{(i)}=0,\right.\nonumber\\
& \phantom{= \frac{1}{2} \min} \quad \left.X=\sum_{i=1}^nX^{(i)}\right)\,.
\end{align}
The equivalence between the two expressions in \eqref{eq:defW1} is proved in \autoref{lem:equiv} of \autoref{app:lemmas}.
\end{defn}

The quantum $W_1$ norm is the maximum norm on $\mathcal{O}_n^T$ such that the difference between each couple of neighboring quantum states has norm at most one.
We define the quantum $W_1$ distance as the distance induced by the quantum $W_1$ norm:
\begin{defn}[Quantum $W_1$ distance]
We define the quantum $W_1$ distance between the quantum states $\rho$ and $\sigma$ of $\mathcal{H}_n$ as
\begin{align}\label{eq:defW1d}
&W_1(\rho,\sigma) = \left\|\rho-\sigma\right\|_{W_1}\nonumber\\
&= \min\left(\sum_{i=1}^n c_i:c_i\ge0,\;\rho - \sigma =\sum_{i=1}^nc_i\left(\rho^{(i)} - \sigma^{(i)}\right),\right.\nonumber\\
& \phantom{= \min} \quad \left.\rho^{(i)},\,\sigma^{(i)}\in\mathcal{S}_n,\;\mathrm{Tr}_i\rho^{(i)} = \mathrm{Tr}_i\sigma^{(i)}\right)\,.
\end{align}
The equivalence between the two expressions in \eqref{eq:defW1d} can be proved along the same lines of \autoref{lem:equiv} of \autoref{app:lemmas}.
\end{defn}

For the sake of a simpler notation, we state all our results in terms of the quantum $W_1$ norm.
Their counterparts for the quantum $W_1$ distance trivially follow.

\section{Properties of the quantum \texorpdfstring{$W_1$}{W1} distance}\label{sec:prop}

\subsection{Relation with the trace distance}\label{sec:tr}
The following \autoref{prop:1T} states that the quantum $W_1$ norm keeps the same upper and lower bounds in terms of the trace norm as its classical counterpart.

\begin{prop}[Relation with the trace norm]\label{prop:1T}
For any $X\in\mathcal{O}_n^T$,
\begin{equation}\label{eq:1T}
\frac{1}{2}\left\|X\right\|_1 \le \left\|X\right\|_{W_1} \le \frac{n}{2}\left\|X\right\|_1\,.
\end{equation}
Moreover, if $\mathrm{Tr}_iX = 0$ for some $i\in[n]$, and in particular if $n=1$,
\begin{equation}\label{eq:1Ti}
\left\|X\right\|_{W_1} = \frac{1}{2}\left\|X\right\|_1\,,
\end{equation}
\emph{i.e.}, for any $\rho,\,\sigma\in\mathcal{S}_n$ such that $\mathrm{Tr}_i\rho = \mathrm{Tr}_i\sigma$ for some $i\in[n]$,
\begin{equation}
\left\|\rho - \sigma\right\|_{W_1} = \frac{1}{2}\left\|\rho - \sigma\right\|_1\,.
\end{equation}
\begin{proof}
On the one hand, let $X^{(1)},\,\ldots,\,X^{(n)}$ be as in \eqref{eq:defW1}.
We have
\begin{equation}
\left\|X\right\|_1 \le \sum_{i=1}^n\left\|X^{(i)}\right\|_1\,,
\end{equation}
therefore
\begin{equation}
\frac{1}{2}\left\|X\right\|_1 \le \left\|X\right\|_{W_1}\,.
\end{equation}
On the other hand, let
\begin{equation}
X = X^+ - X^-\,,
\end{equation}
where $X^+$ and $X^-$ are positive semidefinite with orthogonal supports and satisfy
\begin{equation}
\mathrm{Tr}\,X^{\pm} = \frac{1}{2}\left\|X\right\|_1\,.
\end{equation}
We can choose in \eqref{eq:defW1}
\begin{align}
X^{(i)} &= \frac{2}{\left\|X\right\|_1}\left(\mathrm{Tr}_{i\ldots n}X^-\otimes \mathrm{Tr}_{1\ldots i-1}X^+\right.\nonumber\\
& \phantom{=\frac{2}{\left\|X\right\|_1}} \quad \left. - \mathrm{Tr}_{i+1\ldots n}X^-\otimes \mathrm{Tr}_{1\ldots i}X^+\right)\,,
\end{align}
such that
\begin{equation}
\left\|X^{(i)}\right\|_1 \le \left\|X\right\|_1\,,
\end{equation}
therefore
\begin{equation}
\left\|X\right\|_{W_1} \le \frac{1}{2}\sum_{i=1}^n\left\|X^{(i)}\right\|_1 \le \frac{n}{2}\left\|X\right\|_1\,,
\end{equation}
and the claim \eqref{eq:1T} follows.

Let us now assume that $\mathrm{Tr}_iX=0$.
On the one hand, we have already proved that
\begin{equation}
\frac{1}{2}\left\|X\right\|_1 \le \left\|X\right\|_{W_1}\,.
\end{equation}
On the other hand, choosing in \eqref{eq:defW1}
\begin{align}
X^{(i)} &= X\,,\nonumber\\
X^{(1)} &= \ldots = X^{(i-1)} = X^{(i+1)} = \ldots = X^{(n)} = 0\,,
\end{align}
we get
\begin{equation}
\left\|X\right\|_{W_1} \le \frac{1}{2}\left\|X\right\|_1\,,
\end{equation}
and the claim \eqref{eq:1Ti} follows.
\end{proof}
\end{prop}

\subsection{Symmetries}\label{sec:symm}
The classical $W_1$ distance on probability distributions on $[d]^n$ is invariant with respect to permutations of the $n$ subsystems and to permutations of the $d$ elements of one subsystem.
The following \autoref{prop:LU} states that the quantum $W_1$ norm keeps all the symmetries of the classical case, and the permutations of the $d$ elements of one subsystem get enhanced to unitary operations acting on one qudit.
\begin{prop}[Symmetries of the quantum $W_1$ norm]\label{prop:LU}
The quantum $W_1$ norm is invariant with respect to permutations of the qudits and unitary operations acting on one qudit, and non-increasing with respect to quantum channels acting on one qudit.
\begin{proof}
The claim follows since all the transformations above send $\mathcal{N}_n$ to itself.
\end{proof}
\end{prop}

\subsection{Tensorization}\label{sec:tens}
In the following \autoref{prop:tens}, we prove that the quantum $W_1$ distance is additive with respect to the tensor product as its classical counterpart.
This property is fundamental for distortion measures in rate distortion theory \cite[Chapter 5]{gray2013entropy}, and it is not satisfied by the trace distance.
On the other hand, the quantum relative entropy and the logarithm of the inverse of the quantum fidelity are additive, but they are not proper distances since they do not satisfy the triangle inequality.

\begin{prop}[Tensorization]\label{prop:tens}
For any $X\in\mathcal{O}_{m+n}^T$,
\begin{equation}\label{eq:WT}
\left\|X\right\|_{W_1} \ge \left\|\mathrm{Tr}_{m+1\ldots m+n}X\right\|_{W_1} + \left\|\mathrm{Tr}_{1\ldots m}X\right\|_{W_1}\,,
\end{equation}
and for any $\rho,\,\sigma\in\mathcal{S}_{m+n}$,
\begin{align}
\left\|\rho - \sigma\right\|_{W_1} &\ge \left\|\rho_{1\ldots m}- \sigma_{1\ldots m}\right\|_{W_1}\nonumber\\
&\phantom{\ge} + \left\|\rho_{m+1\ldots m+n} - \sigma_{m+1\ldots m+n}\right\|_{W_1}\,.
\end{align}
Moreover, for any $\rho',\,\sigma'\in\mathcal{S}_{m}$ and any $\rho'',\,\sigma''\in\mathcal{S}_{n}$,
\begin{equation}\label{eq:WT2}
\left\|\rho'\otimes\rho'' - \sigma'\otimes\sigma''\right\|_{W_1} = \left\|\rho' - \sigma'\right\|_{W_1} + \left\|\rho'' - \sigma''\right\|_{W_1}\,.
\end{equation}
\begin{proof}
Let $X^{(1)},\,\ldots,\,X^{(m+n)}\in\mathcal{O}_{m+n}^T$ be such that
\begin{equation}
\mathrm{Tr}_iX^{(i)}=0\quad\forall\,i\in[m+n]\,,\qquad X=\sum_{i=1}^{m+n}X^{(i)}\,.
\end{equation}
We have $\mathrm{Tr}_{m+1\ldots m+n}X^{(i)}=0$ for any $i=m+1,\,\ldots,\,m+n$ and $\mathrm{Tr}_{1\ldots m}X^{(i)}=0$ for any $i\in[m]$, therefore
\begin{align}
\mathrm{Tr}_{m+1\ldots m+n}X &= \sum_{i=1}^{m}\mathrm{Tr}_{m+1\ldots m+n}X^{(i)}\,,\nonumber\\
\mathrm{Tr}_{1\ldots m}X &= \sum_{i=m+1}^{m+n}\mathrm{Tr}_{1\ldots m}X^{(i)}\,,
\end{align}
then
\begin{align}
&\left\|\mathrm{Tr}_{m+1\ldots m+n}X\right\|_{W_1} + \left\|\mathrm{Tr}_{1\ldots m}X\right\|_{W_1}\nonumber\\
&\le \frac{1}{2}\sum_{i=1}^{m}\left\|\mathrm{Tr}_{m+1\ldots m+n}X^{(i)}\right\|_1 + \frac{1}{2}\sum_{i=m+1}^{m+n}\left\|\mathrm{Tr}_{1\ldots m}X^{(i)}\right\|_1\nonumber\\
&\le \frac{1}{2}\sum_{i=1}^{m+n}\left\|X^{(i)}\right\|_1\,,
\end{align}
and the claim \eqref{eq:WT} follows.

On the one hand, we have from \eqref{eq:WT}
\begin{equation}
\left\|\rho'\otimes\rho'' - \sigma'\otimes\sigma''\right\|_{W_1} \ge \left\|\rho' - \sigma'\right\|_{W_1} + \left\|\rho'' - \sigma''\right\|_{W_1}\,.
\end{equation}
On the other hand, we get with the help of \autoref{lem:XOY} of \autoref{app:lemmas}
\begin{align}
&\left\|\rho'\otimes\rho'' - \sigma'\otimes\sigma''\right\|_{W_1}\nonumber\\
&\le \left\|\left(\rho'-\sigma'\right)\otimes\rho''\right\|_{W_1} + \left\|\sigma'\otimes\left(\rho''-\sigma''\right)\right\|_{W_1}\nonumber\\
&\le \left\|\rho'-\sigma'\right\|_{W_1} + \left\|\rho''-\sigma''\right\|_{W_1}\,,
\end{align}
and the claim \eqref{eq:WT2} follows.
\end{proof}
\end{prop}
\begin{cor}\label{cor:add}
For any $\rho,\,\sigma\in\mathcal{S}_n$,
\begin{equation}\label{eq:tproduct}
\left\|\rho - \sigma\right\|_{W_1} \ge \frac{1}{2}\sum_{i=1}^n\left\|\rho_i - \sigma_i\right\|_1\,,
\end{equation}
and equality holds whenever both $\rho$ and $\sigma$ are product states.
\end{cor}

\subsection{Local operations}\label{sec:marg}
The quantum $W_1$ distance between two quantum states that coincide after discarding one qudit is at most one.
In the following \autoref{prop:mar}, we consider the case of quantum states that coincide after discarding $k$ qudits, and we prove that their distance is at most $2k$.
\begin{prop}\label{prop:mar}
Let $\mathcal{I}\subseteq[n]$, and let $X\in\mathcal{O}_n^T$ such that $\mathrm{Tr}_{\mathcal{I}}X=0$.
Then,
\begin{equation}
\left\|X\right\|_{W_1} \le \left|\mathcal{I}\right|\frac{d^2-1}{d^2}\left\|X\right\|_1\,,
\end{equation}
and for any $\rho,\,\sigma\in\mathcal{S}_n$ such that $\mathrm{Tr}_\mathcal{I}\rho = \mathrm{Tr}_\mathcal{I}\sigma$,
\begin{equation}
\left\|\rho - \sigma\right\|_{W_1} \le \left|\mathcal{I}\right|\frac{d^2-1}{d^2}\left\|\rho - \sigma\right\|_1\,.
\end{equation}
\begin{proof}
Without loss of generality, we can assume that $\mathcal{I}=[k]$ for some $k\in[n]$.
For any $i\in[k]$, let
\begin{equation}
X^{(i)} = \frac{\mathbb{I}_d^{\otimes\left(i-1\right)}}{d^{i-1}}\otimes \mathrm{Tr}_{1\ldots i-1}X - \frac{\mathbb{I}_d^{\otimes i}}{d^{i}}\otimes \mathrm{Tr}_{1\ldots i}X\,,
\end{equation}
such that
\begin{equation}
\mathrm{Tr}_iX^{(i)}=0\,,\qquad X = \sum_{i=1}^k X^{(i)}\,.
\end{equation}
We have with the help of \autoref{lem:E} of \autoref{app:lemmas}
\begin{align}
\left\|X\right\|_{W_1} &\le \frac{1}{2}\sum_{i=1}^k\left\|X^{(i)}\right\|_1\nonumber\\
&\le \frac{d^2-1}{d^2}\sum_{i=1}^k\left\|\mathrm{Tr}_{1\ldots i-1}X\right\|_1 \le \left|\mathcal{I}\right|\frac{d^2-1}{d^2}\left\|X\right\|_1\,,
\end{align}
and the claim follows.
\end{proof}
\end{prop}

An important consequence of \autoref{prop:mar} is that the $W_1$ distance is continuous with respect to local operations, in the sense that any operation performed on $k$ qudits can displace the initial quantum state by at most $2k$ in the distance:
\begin{cor}\label{prop:k}
Let $\Phi$ be a quantum channel on $\mathcal{H}_n$ that acts on at most $k$ qudits.
Then, for any $\rho\in\mathcal{S}_n$,
\begin{equation}
\left\|\Phi(\rho) - \rho\right\|_{W_1} \le 2\,k\,\frac{d^2-1}{d^2}\,.
\end{equation}
\begin{proof}
Let $\mathcal{I}\subseteq[n]$ be the set of qudits on which $\Phi$ acts.
Then, $\mathrm{Tr}_\mathcal{I}\left[\Phi(\rho)-\rho\right]=0$, and the claim follows from \autoref{prop:mar}.
\end{proof}
\end{cor}

\subsection{Recovery of the classical \texorpdfstring{$W_1$}{W1} distance}\label{sec:class}
The following \autoref{prop:cl} states that for quantum states diagonal in the canonical basis, the quantum $W_1$ distance recovers the classical $W_1$ distance.

\begin{prop}\label{prop:cl}
Let $p,\,q\in\mathcal{P}_n$, and let
\begin{equation}\label{eq:defpq}
\rho = \sum_{x\in[d]^n}p(x)\,|x\rangle\langle x|\,,\qquad \sigma = \sum_{y\in[d]^n}q(y)\,|y\rangle\langle y|\,.
\end{equation}
Then,
\begin{equation}\label{eq:CQ}
\left\|\rho-\sigma\right\|_{W_1} = W_1(p,q)\,.
\end{equation}
In particular, the quantum $W_1$ distance between vectors of the canonical basis coincides with the Hamming distance:
\begin{equation}\label{eq:WH}
\left\||x\rangle\langle x| - |y\rangle\langle y|\right\|_{W_1} = h(x,y)\,,\qquad x,\,y\in[d]^n\,.
\end{equation}
\begin{proof}
Let $x,\,y\in[d]^n$.
We get from \autoref{cor:add}
\begin{align}
\left\||x\rangle\langle x| - |y\rangle\langle y|\right\|_{W_1} &= \frac{1}{2}\sum_{i=1}^n\left\||x_i\rangle\langle x_i| - |y_i\rangle\langle y_i|\right\|_1\nonumber\\
&= h(x,y)\,,
\end{align}
and the claim \eqref{eq:WH} follows.

On the one hand, let $\pi\in\mathcal{C}(p,q)$.
We have
\begin{align}
\left\|\rho-\sigma\right\|_{W_1} &= \left\|\sum_{x,\,y\in[d]^n}\pi(x,y)\left(|x\rangle\langle x| - |y\rangle\langle y|\right)\right\|_{W_1}\nonumber\\
&\le \sum_{x,\,y\in[d]^n}\pi(x,y)\left\||x\rangle\langle x| - |y\rangle\langle y|\right\|_{W_1}\nonumber\\
&= \sum_{x,\,y\in[d]^n}h(x,y)\,\pi(x,y)\,,
\end{align}
therefore
\begin{equation}
\left\|\rho-\sigma\right\|_{W_1} \le W_1(p,q)\,.
\end{equation}
On the other hand, there exist a probability distribution $r$ on $[n]$ and quantum states $\rho^{(1)},\,\sigma^{(1)},\,\ldots,\,\rho^{(n)},\,\sigma^{(n)}\in\mathcal{S}_n$ such that
\begin{align}
\mathrm{Tr}_i\rho^{(i)} &= \mathrm{Tr}_i\sigma^{(i)}\quad\forall\,i\in[n]\,,\nonumber\\
\rho - \sigma &= \left\|\rho - \sigma\right\|_{W_1}\sum_{i=1}^n r_i\left(\rho^{(i)} - \sigma^{(i)}\right)\,.
\end{align}
We can assume that each $\rho^{(i)}$ and each $\sigma^{(i)}$ is diagonal in the canonical basis.
Let $p^{(1)},\,\ldots,\,p^{(n)}$ and $q^{(1)},\,\ldots,\,q^{(n)}$ be the associated probability distributions on $[d]^n$, such that
\begin{equation}
p -q = \left\|\rho - \sigma\right\|_{W_1}\sum_{i=1}^n r_i\left(p^{(i)} - q^{(i)}\right)\,.
\end{equation}
Since also the classical $W_1$ distance is induced by a norm, we have from \autoref{lem:nc} of \autoref{app:lemmas}
\begin{equation}
W_1(p,q) \le \left\|\rho - \sigma\right\|_{W_1}\sum_{i=1}^n r_i\,W_1\left(p^{(i)},\,q^{(i)}\right) \le \left\|\rho - \sigma\right\|_{W_1}\,,
\end{equation}
and the claim \eqref{eq:CQ} follows.
\end{proof}
\end{prop}

\section{The quantum Lipschitz constant and the dual formulation of the quantum \texorpdfstring{$W_1$}{W1} distance}\label{sec:dual}
The classical $W_1$ distance between the probability distributions $p$ and $q$ on the metric space $\mathcal{X}$ admits a dual formulation as maximum difference between the expectation values of a Lipschitz function on $p$ and $q$:
\begin{align}
&W_1(p,q) =\nonumber\\
&\max\left(\sum_{x\in\mathcal{X}}f(x)\left(p(x) - q(x)\right):f\in\mathbb{R}^{\mathcal{X}},\,\left\|f\right\|_L\le1\right)\,,
\end{align}
where
\begin{equation}
\left\|f\right\|_L = \max_{x\neq y\in\mathcal{X}}\frac{\left|f(x) - f(y)\right|}{D(x,y)}
\end{equation}
is the Lipschitz constant of $f$, and $D$ is the distance on $\mathcal{X}$.
This dual formulation makes the computation of the classical $W_1$ distance a semidefinite program (actually, the same holds for all the $W_\alpha$ distances).

We prove in the following that the computation of the quantum $W_1$ norm is also a semidefinite program.
First, we need to define a quantum generalization of the Lipschitz constant:
\begin{defn}[Quantum Lipschitz constant]
We define the quantum Lipschitz constant of $H\in\mathcal{O}_n$ as the dual norm of the quantum $W_1$ norm on $\mathcal{O}_n^T$:
\begin{align}
&\left\|H\right\|_L = \max\left(\mathrm{Tr}\left[H\,X\right]:X\in\mathcal{O}_n^T,\,\left\|X\right\|_{W_1}\le1\right)\nonumber\\
&= \max\left(\mathrm{Tr}\left[H\,X\right]:X\in\mathcal{N}_n\right)\nonumber\\
&= \max_{i\in[n]}\left(\max\left(\mathrm{Tr}\left[H\left(\rho-\sigma\right)\right]:\rho,\,\sigma\in\mathcal{S}_n,\;\mathrm{Tr}_i\rho = \mathrm{Tr}_i\sigma\right)\right)\,.
\end{align}
\end{defn}
The quantum Lipschitz constant recovers the classical Lipschitz constant for operators diagonal in the canonical basis:

\begin{prop}
Let $f:[d]^n\to\mathbb{R}$, and let
\begin{equation}
F = \sum_{x\in[d]^n}f(x)\,|x\rangle\langle x|\,.
\end{equation}
Then,
\begin{equation}
\left\|F\right\|_L = \left\|f\right\|_L\,.
\end{equation}
\begin{proof}
Let $\mathcal{D}$ be the quantum channel on $\mathbb{C}^d$ that dephases the input state in the canonical basis:
\begin{equation}
\mathcal{D}(X) = \sum_{i=1}^d\langle i|X|i\rangle\,|i\rangle\langle i|\,,\qquad X\in\mathcal{S}_1\,.
\end{equation}
From \autoref{prop:LU}, we have for any $X\in\mathcal{O}_n^T$
\begin{equation}
\left\|\mathcal{D}^{\otimes n}(X)\right\|_{W_1} \le \left\|X\right\|_{W_1}\,.
\end{equation}
We then have with the help of \autoref{prop:cl}
\begin{align}
&\left\|F\right\|_L = \max\left(\sum_{x\in[d]^n}f(x)\left(\langle x|\rho|x\rangle - \langle x|\sigma|x\rangle\right):\right.\nonumber\\
& \phantom{\left\|F\right\|_L = \max} \quad \left.\rho,\,\sigma\in\mathcal{S}_n\,,\quad\left\|\rho-\sigma\right\|_{W_1}\le1\right)\nonumber\\
&= \max\left(\sum_{x\in[d]^n}f(x)\left(\langle x|\mathcal{D}^{\otimes n}(\rho)|x\rangle - \langle x|\mathcal{D}^{\otimes n}(\sigma)|x\rangle\right):\right.\nonumber\\
&\phantom{=\max}\quad\left.\rho,\,\sigma\in\mathcal{S}_n\,,\quad\left\|\mathcal{D}^{\otimes n}(\rho-\sigma)\right\|_{W_1}\le1\right)\nonumber\\
&=\max\left(\sum_{x\in[d]^n}f(x)\left(p(x) - q(x)\right):\right.\nonumber\\
&\phantom{=\max}\quad\left. p,\,q\in\mathcal{P}_n\,,\quad W_1(p,q)\le1\right)\nonumber\\
&= \left\|f\right\|_L \,,
\end{align}
and the claim follows.
\end{proof}
\end{prop}

\autoref{prop:L} of \autoref{app:a} provides an estimate of the quantum Lipschitz constant up to multiplicative error $\sqrt{2}$ that does not require any optimization.
The following \autoref{prop:dualL} provides a dual formulation of the quantum Lipschitz constant:

\begin{prop}\label{prop:dualL}
For any $H\in\mathcal{O}_n$,
\begin{equation}
\left\|H\right\|_L = 2\max_{i\in[n]}\min_{H^{(i)}\in\mathcal{O}_{n-1}}\left\|H - \mathbb{I}_d^{(i)}\otimes H^{(i)}\right\|_\infty\,,
\end{equation}
where for any $i\in[n]$, $\mathbb{I}_d^{(i)}$ is the identity operator on the $i$-th qudit and $H^{(i)}$ does not act on the $i$-th qudit.
\begin{proof}
It is sufficient to prove that
\begin{align}
&\max\left(\mathrm{Tr}\left[H\left(\rho-\sigma\right)\right]:\rho,\,\sigma\in\mathcal{S}_n,\,\mathrm{Tr}_1\rho = \mathrm{Tr}_1\sigma\right)\nonumber\\
& = 2\min_{K\in\mathcal{O}_{n-1}}\left\|H - \mathbb{I}_d\otimes K\right\|_\infty\,.
\end{align}
Let $\Phi:\mathbb{R}\times\mathcal{O}_{n-1}\to\mathcal{O}_n^2$ be given by
\begin{align}
&\Phi(t,K) = \left(t\,\mathbb{I}_d^{\otimes n} + \mathbb{I}_d\otimes K\,,\;t\,\mathbb{I}_d^{\otimes n} - \mathbb{I}_d\otimes K\right)\,,\nonumber\\
& t\in\mathbb{R}\,,\;K\in\mathcal{O}_{n-1}\,,
\end{align}
such that
\begin{align}
&2\min_{K\in\mathcal{O}_{n-1}}\left\|H - \mathbb{I}_d\otimes K\right\|_\infty = 2\min\left(t\in\mathbb{R}:\exists \, K\in\mathcal{O}_{n-1}:\right.\nonumber\\
&\left.\Phi(t,K) - \left(H,\,-H\right) \in\left(\mathcal{O}_n^+\right)^2\right)
\end{align}
is a semidefinite program with dual program
\begin{align}
&\max\left(\mathrm{Tr}\left[H\left(\alpha-\beta\right)
\right]:\alpha,\,\beta\in\mathcal{O}_n^+\,,\;\Phi^\dag(\alpha,\beta)=\left(2,0\right)\right)\nonumber\\
&= \max\left(\mathrm{Tr}\left[H\left(\rho - \sigma\right)\right]:\rho,\,\sigma\in\mathcal{S}_n\,,\;\mathrm{Tr}_1\rho = \mathrm{Tr}_1\sigma\right)\,.
\end{align}
$\left(\mathcal{O}_n^+\right)^2$ and $\mathbb{R}\times\mathcal{O}_{n-1}$ are both convex cones.
Moreover, for any $t>\left\|H\right\|_\infty$ we have
\begin{equation}
\Phi(t,0) - \left(H,\,-H\right) = \left(t\,\mathbb{I}_d^{\otimes n} - H,\,t\,\mathbb{I}_d^{\otimes n} + H\right)\in\mathrm{int}\,\left(\mathcal{O}_n^+\right)^2\,.
\end{equation}
Therefore, from \cite[Corollary 5.3.6]{borwein2013convex} there is no duality gap, and the claim follows.
\end{proof}
\end{prop}

In finite dimension, the dual of the dual norm always coincides with the original norm.
Therefore, the quantum $W_1$ norm is the dual norm of the quantum Lipschitz constant.
Thanks to \autoref{prop:dualL}, this dual formulation of the quantum $W_1$ norm is the dual program of the semidefinite program \eqref{eq:defW1}:
\begin{prop}[Duality]\label{prop:dual}
The optimization problem \eqref{eq:defW1} is a semidefinite program with the following dual program: for any $X\in\mathcal{O}_n^T$,
\begin{align}
&\left\|X\right\|_{W_1} = \max\left(\mathrm{Tr}\left[H\,X\right]:H\in\mathcal{O}_n,\,\left\|H\right\|_L\le1\right)\nonumber\\
&=\max\left(\mathrm{Tr}\left[H\,X\right]:H\in\mathcal{O}_n:\forall\,i\in[n]\;\exists\,H^{(i)}\in\mathcal{O}_{n-1}:\right.\nonumber\\
&\phantom{=\max}\quad \left. \left\|H - \mathbb{I}_d^{(i)}\otimes H^{(i)}\right\|_\infty\le\frac{1}{2}\right)\,.
\end{align}
\end{prop}

\section{\texorpdfstring{$W_1$}{W1} continuity of the von Neumann entropy}\label{sec:ent}
The von Neumann entropy of a quantum state \cite{nielsen2010quantum,wilde2017quantum,holevo2019quantum}
\begin{equation}
S(\rho) = -\mathrm{Tr}\left[\rho\ln\rho\right]\,,\qquad\rho\in\mathcal{S}_n
\end{equation}
quantifies the amount of uncertainty contained in the state and plays a key role in quantum information theory.
The von Neumann entropy is not sensitive to operations performed on a small subsystem: From \autoref{lem:S} of \autoref{app:lemmas}, any operation performed on $k$ qudits can change the entropy of the state by at most $2k\ln d$.
Since already an operation performed on one qudit can generate a quantum state orthogonal to the initial state, this robustness of the von Neumann entropy cannot be captured by any unitarily invariant distinguishability measure, such as the trace distance, the quantum fidelity or the quantum relative entropy.
The situation for the proposed quantum $W_1$ distance is radically different, since it is robust with respect to local perturbations.

In the classical case, the $W_1$ distance provides the following continuity bound for the Shannon entropy:
\begin{thm*}[$W_1$ continuity of the Shannon entropy {\cite[Proposition 8]{polyanskiy2016wasserstein}}]
For any $p,\,q\in\mathcal{P}_n$,
\begin{equation}\label{eq:SWc}
\left|S(p) - S(q)\right|\le n\,h_2\left(\frac{W_1(p,q)}{n}\right) + W_1(p,q)\ln\left(d-1\right)\,,
\end{equation}
where $h_2$ is the binary entropy function
\begin{equation}
h_2(x) = -x\ln x -\left(1-x\right)\ln\left(1-x\right)\,,\qquad 0\le x \le 1\,.
\end{equation}
\begin{proof}
The proof is based on couplings.
For the sake of completeness, we report it in \autoref{app:entc}.
\end{proof}
\end{thm*}

A natural question is whether the continuity bound \eqref{eq:SWc} still holds without any modification for the quantum $W_1$ distance.
The answer is negative.
Indeed, the right-hand side of \eqref{eq:SWc} has a unique maximum equal to $n\ln d$ achieved at $W_1(p,q) = n\left(d-1\right)/d$.
Since $n\ln d$ is the entropy of the maximally mixed state of $\mathcal{H}_n$, the continuity bound \eqref{eq:SWc} would imply that the $W_1$ distance between the maximally mixed state and any pure state is equal to $n\left(d-1\right)/d$.
However, if $\gamma$ is a maximally entangled state acting on $\left(\mathbb{C}^d\right)^{\otimes2}$, from \autoref{lem:EPR} of \autoref{app:lemmas} for any even $n$ we have
\begin{equation}
\left\|\gamma^{\otimes\frac{n}{2}} - \frac{\mathbb{I}_d^{\otimes n}}{d^n}\right\|_{W_1} = \frac{n}{2}\,\frac{d^2-1}{d^2} < n\,\frac{d-1}{d}\,,
\end{equation}
hence the continuity bound \eqref{eq:SWc} cannot hold without modifications in the quantum setting.

Nonetheless, the von Neumann entropy has good continuity properties with respect to the quantum $W_1$ distance.
Indeed, the von Neumann entropy satisfies the following continuity bound, which is equivalent to the classical bound \eqref{eq:SWc} up to a factor $\ln n$:
\begin{thm}[$W_1$ continuity of the von Neumann entropy]\label{thm:ent}
For any $\rho,\,\sigma\in\mathcal{S}_n$,
\begin{equation}\label{eq:entb}
\left|S(\rho) - S(\sigma)\right| \le g\left(\left\|\rho-\sigma\right\|_{W_1}\right) + \left\|\rho-\sigma\right\|_{W_1}\ln\left(d^2\,n\right)\,,
\end{equation}
where for any $t\ge0$
\begin{equation}
g(t) = \left(t+1\right)\ln\left(t+1\right) - t\ln t\,.
\end{equation}
\begin{proof}
Let
\begin{equation}
t = \left\|\rho-\sigma\right\|_{W_1}\,.
\end{equation}
There exist a probability distribution $p$ on $[n]$ and quantum states $\sigma^{(1)},\,\rho^{(1)},\,\ldots,\,\sigma^{(n)},\,\rho^{(n)}\in\mathcal{S}_n$ such that
\begin{align}
\mathrm{Tr}_i\sigma^{(i)} &= \mathrm{Tr}_i\rho^{(i)}\quad\forall\,i\in[n]\,,\nonumber\\
\rho - \sigma &= t\sum_{i=1}^n p_i\left(\rho^{(i)} - \sigma^{(i)}\right)\,.
\end{align}
Let $q$ be the probability distribution on $\left\{0,\,\ldots,\,n\right\}$ given by
\begin{equation}
q_0 = \frac{1}{t+1}\,,\qquad q_i = \frac{t}{t+1}\,p_i\,,\quad i\in[n]\,,
\end{equation}
such that
\begin{equation}
q_0\,\rho + \sum_{i=1}^n q_i\,\sigma^{(i)} = q_0\,\sigma + \sum_{i=1}^n q_i\,\rho^{(i)} = \tau\in\mathcal{S}_n\,.
\end{equation}
We have
\begin{equation}\label{eq:Sq}
S(q) = h_2(q_0) + \left(1-q_0\right)S(p) \le h_2(q_0) + \left(1-q_0\right)\ln n\,.
\end{equation}
Moreover, \autoref{lem:S} of \autoref{app:lemmas} implies for any $i\in[n]$
\begin{equation}\label{eq:Si}
S\left(\rho^{(i)}\right) - S\left(\sigma^{(i)}\right) \le 2\ln d\,.
\end{equation}
On the one hand, we have from the concavity of the entropy
\begin{equation}\label{eq:tauge}
S(\tau) \ge q_0\,S(\rho) + \sum_{i=1}^n q_i\,S\left(\sigma^{(i)}\right)\,.
\end{equation}
On the other hand, we have
\begin{equation}\label{eq:taule}
S(\tau) \le q_0\,S(\sigma) + \sum_{i=1}^n q_i\,S\left(\rho^{(i)}\right) + S(q)\,.
\end{equation}
Putting together \eqref{eq:tauge}, \eqref{eq:taule}, \eqref{eq:Si} and \eqref{eq:Sq} we get
\begin{align}
S(\rho) - S(\sigma) &\le \frac{1}{q_0}\left(\sum_{i=1}^n q_i\left(S\left(\rho^{(i)}\right) - S\left(\sigma^{(i)}\right)\right) + S(q)\right)\nonumber\\
&\le  \frac{1-q_0}{q_0}\ln\left(d^2\,n\right) + \frac{h_2(q_0)}{q_0}\nonumber\\
&= t\ln\left(d^2\,n\right) + \left(t+1\right)\ln\left(t+1\right) - t\ln t\,,
\end{align}
and the claim follows.
\end{proof}
\end{thm}

\autoref{thm:ent} implies that in the limit of large $n$ with fixed $d$ and for any $\epsilon>0$, if
\begin{equation}
\left\|\rho-\sigma\right\|_{W_1} \le \frac{\epsilon\,n}{\ln\left(d^2\,n\right)}\,,
\end{equation}
then
\begin{equation}
\left|S(\rho) - S(\sigma)\right| \le \epsilon\,n +O(\ln n)\,.
\end{equation}
Since the entropy is intimately linked with entanglement, a fundamental consequence of this result is that the entanglement properties of a quantum state are robust with respect to perturbations in the quantum $W_1$ distance with size $o(n/\ln n)$.
For example, we consider a bipartite quantum system $AB$ with each subsystem consisting of $n$ qudits.
Let $\rho_{AB}$ be a pure quantum state of $AB$ with entanglement entropy and distillable entanglement
\begin{equation}
E_D(\rho_{AB}) = S(\rho_A) = O(n)\,.
\end{equation}
For any perturbation that degrades the quantum state $\rho_{AB}$ to some state $\rho'_{AB}$ such that
\begin{equation}
\left\|\rho_{AB}-\rho'_{AB}\right\|_{W_1} = o\left(\frac{n}{\ln n}\right)\,,
\end{equation}
we have
\begin{equation}
\left|S(\rho_{AB}) - S(\rho'_{AB})\right| = o(n)\,,\qquad \left|S(\rho_B) - S(\rho'_B)\right| = o(n)\,,
\end{equation}
and from \cite[Theorem 3.1]{devetak2005distillation}, the distillable entanglement of $\rho'_{AB}$ is at least
\begin{equation}
E_D(\rho'_{AB}) \ge S(\rho'_B) - S(\rho'_{AB}) = E_D(\rho_{AB}) - o(n)\,.
\end{equation}

\section{Quantum Marton's transportation inequality}\label{sec:Marton}

The quantum relative entropy between two quantum states \cite{nielsen2010quantum,wilde2017quantum,holevo2019quantum}
\begin{equation}
 S(\rho \| \sigma) =  \mathrm{Tr}\left[\rho\left(\ln\rho- \ln \sigma\right)\right]\,,\qquad\rho,\,\sigma \in\mathcal{S}_n\,,
\end{equation}
generalizes the classical Kullback--Leibler divergence. As in the classical case, it is always nonnegative and equal to zero if and only if $\rho = \sigma$.
It can be thought as a distance between quantum states, but it is not symmetric nor it satisfies the triangle inequality.  The quantum Pinsker's inequality \cite[Theorem 11.9.1]{wilde2017quantum}, \cite[Eq. (14.38)]{bengtsson2017geometry}
\begin{equation}\label{eq:qpinsker}
 \left\| \rho - \sigma \right \|_1 \le \sqrt{2\, S( \rho \| \sigma)}
\end{equation}
provides an upper bound for the trace distance in terms of the quantum relative entropy.
In the classical case, an inequality by Marton \cite{marton1986simple} extends Pinsker's inequality to a transportation cost --- information  inequality, by replacing the left hand side with the $W_1$ distance induced by the Hamming distance: if $p$, $q$ are probability distributions on $[d]^n$ and $q$ is a product distribution $q(x)= \prod_{i=1}^n q_i(x_i)$, then
\begin{equation}\label{eq:marton}
W_1( p, q) \le \sqrt{  \frac{ n}{2}\,S( p \| q) }\,.
\end{equation}
Marton's inequality \eqref{eq:marton} improves the classical Pinsker's inequality whenever
\begin{equation}
W_1(p,q)\ge\frac{\sqrt{n}}{2}\left\|p-q\right\|_1\,,
\end{equation}
and was later extended to a larger class of distributions in discrete and continuous settings \cite{marton1996bounding, talagrand1996transportation}. Noncommutative versions of \eqref{eq:marton} and related functional concentration inequalities are proposed in \cite{junge2015noncommutative, rouze2019concentration, osborne2009quantum}, with different quantum generalizations of the Wasserstein distances.
In the following \autoref{thm:marton}, we prove that the proposed quantum $W_1$ distance satisfies the Marton's inequality \eqref{eq:marton}:

\begin{thm}[Quantum Marton's transportation inequality]\label{thm:marton}
For any $\rho, \sigma \in  \mathcal{S}_n$, with $\sigma = \sigma_1 \otimes \ldots \otimes \sigma_n$ product state,
\begin{equation}\label{eq:Qmarton}
\| \rho - \sigma\|_{W_1} \le \sqrt{  \frac{ n}{2}\,S( \rho \| \sigma) }.
\end{equation}
\begin{proof}
As in the proof of \autoref{prop:1T}, we write
\begin{equation}
\rho - \sigma = \sum_{i=1}^n \left( \rho_{1\ldots i} \otimes \sigma_{i+1 \ldots n} - \rho_{1\ldots i-1} \otimes \sigma_{i \ldots n}\right),
\end{equation}
so that
\begin{equation}
 \| \rho - \sigma\|_{W_1} \le \frac{1}{2}\sum_{i=1}^n \left\| \rho_{1\ldots i} \otimes \sigma_{i+1 \ldots n} - \rho_{1\ldots i-1} \otimes \sigma_{i \ldots n}\right\|_1.
\end{equation}
We apply \eqref{eq:qpinsker} for every $i =1$, \ldots, $n$,
\begin{align}
& \left\| \rho_{1\ldots i} \otimes \sigma_{i+1 \ldots n} - \rho_{1\ldots i-1} \otimes \sigma_{i \ldots n}\right\|_1\nonumber\\
&\le \sqrt{2\, S\left( \rho_{1\ldots i} \otimes \sigma_{i+1 \ldots n} \| \rho_{1\ldots i-1} \otimes \sigma_{i \ldots n}\right)}
 \end{align}
and use the concavity of the square root to obtain
\begin{equation}
\| \rho - \sigma\|_{W_1} \le \sqrt{\frac{ n }{2} \sum_{i=1}^n S\left( \rho_{1\ldots i} \otimes \sigma_{i+1 \ldots n} \| \rho_{1\ldots i-1} \otimes \sigma_{i \ldots n}\right) }.
\end{equation}
Using the identity
\begin{align}
&S\left( \rho_{1\ldots i} \otimes \sigma_{i+1 \ldots n} \| \rho_{1\ldots i-1} \otimes \sigma_{i \ldots n}\right)\nonumber\\
& =  S\left( \rho_{1\ldots i} \| \rho_{1\ldots i-1} \otimes \sigma_{i}\right)  \nonumber\\
 & = -S(\rho_{1\ldots i} ) +S(\rho_{1\ldots i-1})- \mathrm{Tr}\left[ \rho_i \log \sigma_i\right]
\end{align}
and telescopic summation, we conclude that
\begin{align}
&\sum_{i=1}^n S\left( \rho_{1\ldots i} \otimes \sigma_{i+1 \ldots n} \| \rho_{1\ldots i-1} \otimes \sigma_{i \ldots n}\right)\nonumber\\
& = -S(\rho) -  \sum_{i=1}^n \mathrm{Tr}\left[ \rho_i \log \sigma_i\right] = -S(\rho) - \mathrm{Tr}\left[\rho \log \sigma \right]\nonumber\\
& = S(\rho \| \sigma)\,,
\end{align}
and the claim follows.
\end{proof}
\end{thm}

\begin{rem}
As in the classical case, the quantum Marton's inequality \eqref{eq:Qmarton} improves the quantum Pinsker's inequality \eqref{eq:qpinsker} whenever
\begin{equation}
\left\|\rho - \sigma\right\|_{W_1} \ge \frac{\sqrt{n}}{2}\left\|\rho - \sigma\right\|_1\,.
\end{equation}
\end{rem}

\section{Quantum Gaussian concentration inequality}\label{sec:conc}
A fundamental consequence of the classical Marton's transportation inequality is Talagrand's inequality \cite{talagrand1996new}, which is a Gaussian measure concentration result for Lipschitz functions.
Talgrand's inequality states that any function that depends smoothly on many independent random variables, but not too much on any of them, must be essentially constant. This is a far-reaching extension of the law of large numbers for sample means of independent random variables, allowing for functions whose dependence on the many variables are quite implicit and computations may not be performed directly.
As illustrated in Refs. \cite{gozlan2010transport, raginsiky2013concentration, boucheron2013concentration}, Talagrand's inequality is a quite general and versatile theoretical tool, with applications ranging from random combinatorial optimization to statistical physics and information theory.

Our first result in the quantum setting is the following quantum Gaussian concentration inequality, which can be thought as an upper bound to the partition function of a quantum Hamiltonian in terms of its quantum Lipschitz constant:

\begin{thm}[Quantum Gaussian concentration inequality]\label{thm:conc}
For any $H\in\mathcal{O}_n$ and any $t\in\mathbb{R}$,
\begin{equation}\label{eq:conc-mgf}
\frac{1}{d^n}\,\mathrm{Tr}\exp\left(t\left(H - \frac{\mathrm{Tr}\,H}{d^n}\,\mathbb{I}\right)\right) \le \exp\frac{n\,t^2\left\|H\right\|_L^2}{8}\,.
\end{equation}
\begin{proof}
Without loss of generality, we can assume that $\mathrm{Tr}\,H = 0$ and $\left\|H\right\|_L=1$, such that the claim becomes
\begin{equation}
\mathrm{Tr}\,\mathrm{e}^{tH} \le d^n\exp\frac{n\,t^2}{8}\,.
\end{equation}
From \autoref{thm:marton} and \autoref{prop:dual}, we have for any $\rho\in\mathcal{S}_n$
\begin{align}\label{eq:martonapp}
S\left(\rho\left\|\frac{\mathbb{I}_d^{\otimes n}}{d^n}\right.\right) &\ge \frac{2}{n}\left\|\rho - \frac{\mathbb{I}_d^{\otimes n}}{d^n}\right\|_{W_1}^2 \ge \frac{2}{n}\left(\mathrm{Tr}\left[H\,\rho\right]\right)^2\nonumber\\
&\ge t\,\mathrm{Tr}\left[H\,\rho\right] - \frac{n\,t^2}{8}\,.
\end{align}
\eqref{eq:martonapp} can be recast as
\begin{equation}
S\left(\rho\left\|\frac{\mathrm{e}^{tH}}{\mathrm{Tr}\,\mathrm{e}^{tH}}\right.\right) - \ln\mathrm{Tr}\,\mathrm{e}^{tH} + n\ln d + \frac{n\,t^2}{8} \ge0\,,
\end{equation}
and the claim follows choosing
\begin{equation}
\rho = \frac{\mathrm{e}^{tH}}{\mathrm{Tr}\,\mathrm{e}^{tH}}\,.
\end{equation}
\end{proof}
\end{thm}

The left-hand side of \eqref{eq:conc-mgf} can be interpreted as the moment generating function of the empirical distribution associated to the spectrum of $H$.
The right-hand side of \eqref{eq:conc-mgf} is the moment generating function of a centered Gaussian distribution with standard deviation $ \sqrt{n} \left\|H\right\|_L/2$.
The inequality \eqref{eq:conc-mgf} implies that the tails of the distribution of the eigenvalues of $H$ decay at least as fast as those of a Gaussian, hence the term ``Gaussian concentration inequality''.
This consequence of \autoref{thm:conc} leads to the following concentration inequality for the spectrum of $H$:

\begin{cor}\label{cor:conc}
Most of the eigenvalues of $H\in\mathcal{O}_n$ lie in an interval with size $O\left(\sqrt{n}\left\|H\right\|_L\right)$, \emph{i.e.}, for any $\delta\ge0$,
\begin{equation}
\dim\left(H \ge \left(\frac{\mathrm{Tr}\,H}{d^n}+\delta\sqrt{n}\left\|H\right\|_L\right)\mathbb{I}\right) \le d^n\,\mathrm{e}^{-2\delta^2}\,,
\end{equation}
where for any $X,\,Y\in\mathcal{O}_n$, $\dim\left(X\ge Y\right)$ denotes the number of nonnegative eigenvalues of $X-Y$.
\begin{proof}
Without loss of generality, we can assume that $\mathrm{Tr}\,H = 0$ and $\left\|H\right\|_L=1$, such that the claim becomes
\begin{equation}
\dim\left(H \ge \delta\sqrt{n}\,\mathbb{I}\right) \le d^n\,\mathrm{e}^{-2\delta^2}\,.
\end{equation}
From \autoref{thm:conc}, we have for any $t\ge0$
\begin{align}
d^n\exp\frac{n\,t^2}{8} \ge \mathrm{Tr}\,\mathrm{e}^{tH} \ge \mathrm{e}^{t\delta\sqrt{n}}\dim\left(H \ge \delta\sqrt{n}\,\mathbb{I}\right)\,,
\end{align}
and the claim follows choosing
\begin{equation}
t = \frac{4\,\delta}{\sqrt{n}}\,.
\end{equation}
\end{proof}
\end{cor}

\autoref{thm:conc} and \autoref{cor:conc} can find application in the field of many-body quantum systems to determine properties of the spectrum of local Hamiltonians, whose quantum Lipschitz constant can be easily controlled:
\begin{prop}\label{prop:lipH}
The quantum Lipschitz constant of a local Hamiltonian is upper bounded by the maximum among the operator norms of the sum of the Hamiltonian terms associated to each qudit.
Formally, let
\begin{equation}\label{eq:localH}
H = \sum_{\mathcal{I}\subseteq[n]}H_\mathcal{I}\,,
\end{equation}
where for every $\mathcal{I}\subseteq[n]$, $H_\mathcal{I}\in\mathcal{O}_n$ has support on the qudits in $\mathcal{I}$ (\emph{e.g.}, if the qudits are arranged in a one-dimensional chain with nearest-neighbors interactions, the only nonzero terms in the sum \eqref{eq:localH} are the $2n-1$ terms associated to the subsets of $[n]$ of the form $\left\{k\right\}$ or $\left\{k,\,k+1\right\}$).
Then,
\begin{equation}
\left\|H\right\|_L \le 2\max_{i\in[n]}\left\|\sum_{\mathcal{I}\subseteq[n]:i\in\mathcal{I}}H_\mathcal{I}\right\|_\infty\,.
\end{equation}
\begin{proof}
Let $X\in\mathcal{N}_n$, and let $i\in[n]$ such that $\mathrm{Tr}_iX=0$.
We have
\begin{equation}
\mathrm{Tr}\left[H\,X\right] = \sum_{\mathcal{I}\subseteq[n]:i\in\mathcal{I}}\mathrm{Tr}\left[H_\mathcal{I}\,X\right] \le 2\left\|\sum_{\mathcal{I}\subseteq[n]:i\in\mathcal{I}}H_\mathcal{I}\right\|_\infty\,,
\end{equation}
and the claim follows.
\end{proof}
\end{prop}

\section{Contraction coefficient}\label{sec:ch}
A fundamental property of the trace distance is that it is contractive with respect to the action of a quantum channel \cite{nielsen2010quantum,wilde2017quantum,holevo2019quantum}, \emph{i.e.}, for any quantum channel $\Phi:\mathcal{O}_n\to\mathcal{O}_m$ and any $\rho,\,\sigma\in\mathcal{S}_n$,
\begin{equation}\label{eq:contr1}
\left\|\Phi(\rho) - \Phi(\sigma)\right\|_1 \le \left\|\rho - \sigma\right\|_1\,.
\end{equation}
The inequality \eqref{eq:contr1} can be sharpened to
\begin{equation}\label{eq:contr1eta}
\left\|\Phi(\rho) - \Phi(\sigma)\right\|_1 \le \eta(\Phi)\left\|\rho - \sigma\right\|_1\,,
\end{equation}
where
\begin{equation}
\eta(\Phi) = \max_{\rho\neq\sigma\in\mathcal{S}_n}\frac{\left\|\Phi(\rho)-\Phi(\sigma)\right\|_1}{\left\|\rho-\sigma\right\|_1}
\end{equation}
is called contraction coefficient of $\Phi$ with respect to the trace distance \cite{reeb2011hilbert,kastoryano2013quantum}, and is strictly smaller than one for any quantum channel with a unique fixed point.

In this section, we explore the contraction properties of the quantum $W_1$ distance.
Since any quantum channel $\Phi$ is trace preserving, it sends $\mathcal{O}_n^T$ to $\mathcal{O}_m^T$.
We denote by
\begin{align}
\left\|\Phi\right\|_{W_1\to W_1} &= \max\left(\left\|\Phi(X)\right\|_{W_1}:X\in\mathcal{O}_n^T,\,\left\|X\right\|_{W_1}\le1\right)\nonumber\\
&= \max_{X\in\mathcal{N}_n}\left\|\Phi(X)\right\|_{W_1}
\end{align}
the norm of $\Phi$ restricted to $\mathcal{O}_n^T$ with respect to the quantum $W_1$ norm, which can also be expressed as
\begin{equation}
\left\|\Phi\right\|_{W_1\to W_1} = \max_{\rho\neq\sigma\in\mathcal{S}_n}\frac{\left\|\Phi(\rho)-\Phi(\sigma)\right\|_{W_1}}{\left\|\rho-\sigma\right\|_{W_1}}\,,
\end{equation}
and is therefore equal to the contraction coefficient of $\Phi$ with respect to the quantum $W_1$ distance.

\subsection{Tensor power channels}
From \autoref{prop:LU}, any quantum operation acting on one qudit cannot expand the quantum $W_1$ distance.
Therefore, for any quantum channel $\Phi$ on $\mathbb{C}^d$, the contraction coefficient of $\Phi^{\otimes n}$ with respect to the quantum $W_1$ distance is at most $1$:
\begin{equation}
\left\|\Phi^{\otimes n}\right\|_{W_1\to W_1} \le 1\,,
\end{equation}
as the contraction coefficient with respect to the trace distance.

Assuming that the output of $\Phi$ is not independent of the input, in the limit $n\to\infty$ the contraction coefficient of $\Phi^{\otimes n}$ with respect to the trace distance is trivial:
\begin{equation}
\lim_{n\to\infty}\eta\left(\Phi^{\otimes n}\right) = 1\,.
\end{equation}
Indeed, for any $\rho,\,\sigma\in\mathcal{S}_1$ such that $\Phi(\rho)\neq\Phi(\sigma)$ we have
\begin{equation}
\lim_{n\to\infty}\left\|\Phi^{\otimes n}\left(\rho^{\otimes n}\right) - \Phi^{\otimes n}\left(\sigma^{\otimes n}\right)\right\|_1 = 2\,.
\end{equation}
For the quantum $W_1$ distance, the situation is radically different.
Indeed, the following \autoref{prop:WWprod} provides a nontrivial upper bound to the contraction coefficient of $\Phi^{\otimes n}$ which does not depend on $n$.
When $\Phi$ is a quantum Markov semigroup, \autoref{prop:WWprod} bounds the worst-case convergence to the equilibrium state.

\begin{prop}\label{prop:WWprod}
Let $\Phi$ be a quantum channel on $\mathbb{C}^d$, let $\omega\in\mathcal{S}_1$ be a fixed point of $\Phi$, and let $\mathcal{E}$ be the quantum channel on $\mathbb{C}^d$ that replaces the input state with $\omega$. Then,
\begin{equation}\label{eq:PhinWW}
\frac{1}{2}\left\|\Phi - \mathcal{E}\right\|_{1\to1} \le \left\|\Phi^{\otimes n}\right\|_{W_1\to W_1} \le \left\|\Phi - \mathcal{E}\right\|_\diamond\,,
\end{equation}
where we recall that for any linear map $\mathcal{F}$ on $\mathcal{O}_1$,
\begin{align}
\left\|\mathcal{F}\right\|_{1\to1} &= \max_{\rho\in\mathcal{S}_1}\left\|\mathcal{F}(\rho)\right\|_1\,,\nonumber\\
\left\|\mathcal{F}\right\|_\diamond &= \max_{\rho\in\mathcal{S}_2}\left\|\left(\mathcal{F}\otimes\mathbb{I}_{\mathcal{O}_1}\right)(\rho)\right\|_1\,.
\end{align}
Therefore, for any $\rho,\,\sigma\in\mathcal{S}_n$,
\begin{equation}
\left\|\Phi^{\otimes n}(\rho) - \Phi^{\otimes n}(\sigma)\right\|_{W_1} \le \left\|\Phi - \mathcal{E}\right\|_\diamond\left\|\rho-\sigma\right\|_{W_1}\,.
\end{equation}
\begin{proof}
Let $X\in\mathcal{N}_n$.
Then, $\mathrm{Tr}_iX=0$ for some $i\in[n]$.
Without loss of generality, we can assume that $i=1$.
Since $\mathrm{Tr}_1\Phi^{\otimes n}(X)=0$, we have from \eqref{eq:1Ti}
\begin{align}
\left\|\Phi^{\otimes n}(X)\right\|_{W_1} &= \frac{1}{2}\left\|\Phi^{\otimes n}(X)\right\|_1 \le \frac{1}{2}\left\|\left(\Phi\otimes\mathbb{I}_{\mathcal{O}_{n-1}}\right)(X)\right\|_1\nonumber\\
&= \frac{1}{2}\left\|\left(\left(\Phi - \mathcal{E}\right)\otimes\mathbb{I}_{\mathcal{O}_{n-1}}\right)(X)\right\|_1\nonumber\\
&\le \frac{1}{2}\left\|\Phi-\mathcal{E}\right\|_\diamond\left\|X\right\|_1 \le \left\|\Phi-\mathcal{E}\right\|_\diamond\,,
\end{align}
where we have also used that $\left(\mathcal{E}\otimes\mathbb{I}_{\mathcal{O}_{n-1}}\right)(X)=0$, therefore
\begin{equation}
\left\|\Phi^{\otimes n}\right\|_{W_1\to W_1} \le \left\|\Phi - \mathcal{E}\right\|_\diamond\,.
\end{equation}
Let $\rho\in\mathcal{S}_1$, and let
\begin{equation}
X = \left(\rho - \omega\right)\otimes\omega^{\otimes\left(n-1\right)}\in\mathcal{N}_n\,.
\end{equation}
We have
\begin{align}
\left\|\Phi^{\otimes n}(X)\right\|_{W_1} &= \frac{1}{2}\left\|\Phi^{\otimes n}(X)\right\|_1 = \frac{1}{2}\left\|\Phi(\rho)-\omega\right\|_1\nonumber\\
&= \frac{1}{2}\left\|\left(\Phi-\mathcal{E}\right)(\rho)\right\|_1\,,
\end{align}
therefore
\begin{equation}
\left\|\Phi^{\otimes n}\right\|_{W_1\to W_1} \ge \frac{1}{2}\left\|\Phi - \mathcal{E}\right\|_{1\to1}\,,
\end{equation}
and the claim follows.
\end{proof}
\end{prop}

We consider the quantum amplitude damping channel as example of application of \autoref{prop:WWprod}:

\begin{ex}[Amplitude damping channel]\label{ex:ad}
Let $d=2$, and for any $0\le p\le1$, let $\Phi_p$ be the quantum amplitude damping channel with decay probability $1-p$ whose action on the Pauli matrices is
\begin{align}
\Phi_p(\mathbb{I}_2) &= \mathbb{I}_2 + \left(1-p\right)\sigma_z\,,\qquad \Phi_p(\sigma_x) = \sqrt{p}\,\sigma_x\,,\nonumber\\
\Phi_p(\sigma_y) &= \sqrt{p}\,\sigma_y\,,\qquad \Phi_p(\sigma_z) = p\,\sigma_z\,.
\end{align}
Then, for any $0\le p\le1/5$,
\begin{equation}\label{eq:W1W1}
\frac{1}{2}\sqrt{\frac{p}{1-p}} \le \left\|\Phi_p^{\otimes n}\right\|_{W_1\to W_1} \le 2\sqrt{\frac{p}{1-p}}\,,
\end{equation}
and for any $\rho,\,\sigma\in\mathcal{S}_n$,
\begin{equation}
\left\|\Phi_p^{\otimes n}(\rho) - \Phi_p^{\otimes n}(\sigma)\right\|_{W_1} \le 2\sqrt{\frac{p}{1-p}}\left\|\rho-\sigma\right\|_{W_1}\,.
\end{equation}
\begin{rem}
For $1/5\le p \le1$, the upper bound of \eqref{eq:W1W1} is trivial since from \autoref{prop:LU}, $\Phi_p^{\otimes n}$ is a contraction for the quantum $W_1$ norm.
\end{rem}
\begin{proof}
The only fixed quantum state of $\Phi_p$ is
\begin{equation}
\omega = \frac{\mathbb{I}_2 + \sigma_z}{2}\,.
\end{equation}
We parameterize a pure state $\rho\in\mathcal{S}_1$ as
\begin{equation}
\rho = \frac{\mathbb{I}_2 + v_x\,\sigma_x + v_y\,\sigma_y + v_z\,\sigma_z}{2}\,,
\end{equation}
where $v$ is a unit vector in $\mathbb{R}^3$.
We have
\begin{align}\label{eq:vz}
&\left\|\Phi_p(\rho) - \omega\right\|_1 = \frac{\sqrt{p}}{2}\left\|v_x\,\sigma_x + v_y\,\sigma_y + \sqrt{p}\left(v_z-1\right)\sigma_z\right\|_1\nonumber\\
&= \sqrt{p\left(v_x^2 + v_y^2 + p\left(1-v_z\right)^2\right)}\nonumber\\
&= \sqrt{p\left(1-v_z\right)\left(1 + p +\left(1-p\right)v_z\right)} \le \sqrt{\frac{p}{1-p}}\,,
\end{align}
where we have used that $v^2=1$.
Let $\mathcal{E}$ be the quantum channel on $\mathbb{C}^2$ that replaces the input state with $\omega$.
We have
\begin{equation}
\left\|\Phi_p - \mathcal{E}\right\|_\diamond \le 2\left\|\Phi_p - \mathcal{E}\right\|_{1\to1} = 2\sqrt{\frac{p}{1-p}}\,,
\end{equation}
and the claim follows from \autoref{prop:WWprod}.
\end{proof}
\end{ex}

We can determine exactly the quantum coefficient of the quantum depolarizing channel with respect to the quantum $W_1$ distance:

\begin{prop}[Quantum depolarizing channel]
Let $\omega\in\mathcal{S}_1$, and for any $0\le p\le1$, let $\mathcal{E}_p$ be the quantum channel on $\mathbb{C}^d$ that is the identity with probability $p$ and replaces the input state with $\omega$ with probability $1-p$:
\begin{equation}
\mathcal{E}_p(X) = p\,X + \left(1-p\right)\omega\,\mathrm{Tr}\,X\,,\qquad X\in\mathcal{O}_1\,.
\end{equation}
Then,
\begin{equation}
\left\|\mathcal{E}_p^{\otimes n}\right\|_{W_1\to W_1} = p\,,
\end{equation}
and for any $\rho,\,\sigma\in\mathcal{S}_n$,
\begin{equation}
\left\|\mathcal{E}_p^{\otimes n}(\rho) - \mathcal{E}_p^{\otimes n}(\sigma)\right\|_{W_1} \le p\left\|\rho-\sigma\right\|_{W_1}\,.
\end{equation}
\begin{proof}
Let $X\in\mathcal{N}_n$, and let $i\in[n]$ be such that $\mathrm{Tr}_iX=0$.
Without loss of generality, we can assume that $i=1$.
We then have from \eqref{eq:1Ti}
\begin{align}\label{eq:ineqch}
&\left\|\mathcal{E}_p^{\otimes n}(X)\right\|_{W_1} = \frac{1}{2}\left\|\mathcal{E}_p^{\otimes n}(X)\right\|_1\nonumber\\
&= \frac{p}{2}\left\|\left(\mathbb{I}_{\mathcal{O}_1}\otimes\mathcal{E}_p^{\otimes\left(n-1\right)}\right)(X)\right\|_1 \le \frac{p}{2}\left\|X\right\|_1 \le p\,,
\end{align}
therefore
\begin{equation}
\left\|\mathcal{E}_p^{\otimes n}\right\|_{W_1\to W_1} \le p\,.
\end{equation}
On the other hand,
\begin{equation}
X = \left(|1\rangle\langle 1| - |2\rangle\langle 2|\right)\otimes \omega^{\otimes\left(n-1\right)}
\end{equation}
achieves equality in \eqref{eq:ineqch}, and the claim follows.
\end{proof}
\end{prop}

\subsection{Shallow quantum circuits}\label{subsec:shallow}
Quantum channels acting on multiple qudits can in general expand the quantum $W_1$ distance.
In the following \autoref{prop:lc}, we prove that the expansion factor is bounded by the size of the light-cones of the input qudits, which can be easily bounded if the channel can be implemented by a shallow local quantum circuit:

\begin{prop}\label{prop:lc}
Let $\Phi:\mathcal{O}_n\to\mathcal{O}_m$ be a quantum channel.
For any $i\in[n]$, let $\mathcal{I}_i\subseteq[m]$ be the light-cone of the $i$-th qudit, \emph{i.e.}, the minimum subset of qudits such that $\mathrm{Tr}_{\mathcal{I}_i}\Phi(X) = 0$ for any $X\in\mathcal{O}_n$ such that $\mathrm{Tr}_iX=0$.
Then,
\begin{equation}
\left\|\Phi\right\|_{W_1\to W_1} \le 2\,\frac{d^2-1}{d^2}\max_{i\in[n]}\left|\mathcal{I}_i\right|\,,
\end{equation}
and for any $\rho,\,\sigma\in\mathcal{S}_n$
\begin{equation}
\left\|\Phi(\rho) - \Phi(\sigma)\right\|_{W_1} \le 2\,\frac{d^2-1}{d^2}\max_{i\in[n]}\left|\mathcal{I}_i\right|\left\|\rho-\sigma\right\|_{W_1}\,.
\end{equation}
\begin{proof}
Let $X\in\mathcal{N}_n$.
Then, $\mathrm{Tr}_iX=0$ for some $i\in[n]$, hence $\mathrm{Tr}_{\mathcal{I}_i}\Phi(X) = 0$.
\autoref{prop:mar} implies
\begin{equation}
\left\|\Phi(X)\right\|_{W_1} \le \left|\mathcal{I}_i\right|\frac{d^2-1}{d^2}\left\|\Phi(X)\right\|_1 \le 2\,\frac{d^2-1}{d^2}\max_{i\in[n]}\left|\mathcal{I}_i\right|\,,
\end{equation}
and the claim follows.
\end{proof}
\end{prop}

\section{Future perspectives}\label{sec:appl}
We have proposed a quantum generalization of the $W_1$ distance which recovers the classical $W_1$ distance as a special case and keeps most of its properties, among which the continuity of the entropy.
In the classical setting, the Wasserstein distances have a huge variety of applications ranging from mathematical analysis to machine learning and information theory.
We expect the proposed quantum $W_1$ distance to be a powerful tool with a broad range of applications in quantum information, quantum computing and quantum machine learning.
We propose a few of them in the following.

\begin{itemize}

    \item {\bf Quantum state estimation}

Estimating a quantum state of $n$ qudits up to $o(1)$ error in the trace distance is a notoriously difficult task, since the number of required copies of the state grows exponentially with $n$ \cite{aaronson2007learnability}.
Requiring instead the quantum $W_1$ distance between the true quantum state and its estimate to be $o(n)$ is a much weaker condition, and the number of required copies can be much smaller.
Therefore, in all the situations where a precision guarantee in terms of the quantum $W_1$ distance is sufficient, employing this distance rather than the trace distance can lead to a significant improvement to the complexity of the estimate.

    \item {\bf Robustness of quantum machine learning}

A fundamental desirable property of classical machine learning algorithms is the robustness with respect to small perturbations in the input \cite{goodfellow2014explaining}, and the same property should be desirable also when the machine learning algorithm is quantum \cite{biamonte2017quantum}.

\emph{Quantum input:} In the scenario with quantum input data, the size of the perturbations in the input has so far been measured with the trace distance or with the quantum fidelity \cite{liu2020vulnerability}, with respect to which any two perfectly distinguishable quantum states are maximally far.
On the contrary, in the classical setting any two different inputs are perfectly distinguishable, and when the input is a bit string the size of the perturbations is measured with the Hamming distance.
Since the proposed quantum $W_1$ distance recovers the Hamming distance for vectors of the canonical basis, it is a perfect candidate to measure the size of the perturbations for quantum algorithms for machine learning with a quantum input.
Therefore, the proposed quantum $W_1$ distance provides a suitable quality factor for the robustness of the quantum algorithms for machine learning.

\emph{Classical input:} In the scenario with classical input data, choosing the right method to encode the input into quantum states is essential in the success of any quantum algorithm for machine learning \cite{biamonte2017quantum,lloyd2020quantum}.
In particular, it is reasonable to require the encoding to be robust with respect to small perturbations of the input. The trivial encoding maps each bit string to the corresponding computational basis state, and is not continuous with respect to any unitarily invariant distance, since any bit flip on the input transforms the quantum state into an orthogonal state.
On the contrary, the trivial encoding is continuous with respect to the proposed quantum $W_1$ distance, since it recovers the Hamming distance  for vectors of the canonical basis.
Therefore, the quantum $W_1$ distance provides a natural measure for the size of the input perturbations and hence for the robustness of the encoding, favoring encodings that map classical inputs with small Hamming distance into quantum states with small quantum $W_1$ distance.

  \item {\bf Quantum Generative Adversarial Networks}

In analogy to classical GANs, quantum GANs \cite{lloyd2018quantum} are a paradigm for quantum machine learning where a generator tries to produce quantum samples as close as possible to some true quantum data, and a discriminator tries to discriminate the generated from the true data.
For classical GANs, the Wasserstein distances have turned out to be the best candidate for the loss function, since they solve the problem of the vanishing gradient in the training that plagued the GANs trained with the total variation distance or with the Jensen--Shannon divergence \cite{arjovsky2017wasserstein}.
For this reason, quantum Wasserstein distances have been proposed as cost function for the quantum GANs \cite{chakrabarti2019quantum,becker2021quantum}.
The proposed quantum $W_1$ distance recovers the classical $W_1$ distance for states diagonal in the canonical basis and satisfies most of its properties, and is therefore a good candidate for the loss function of the quantum GANs.

  \item {\bf Quantum rate distortion theory}

Rate-distortion theory addresses the problem of determining the maximum compression rate of a signal if a certain level of distortion in the recovered signal is allowed \cite{gray2013entropy}.
The measure employed to quantify the distortion plays a fundamental role, and for a discrete alphabet the most prominent distortion measure is the Hamming distance.
Rate-distortion theory has been extended to the quantum setting in the iid regime \cite{barnum2000quantum,devetak2001quantum,devetak2002quantum,chen2008entanglement, datta2012quantum,datta2013quantum,wilde2013quantum,salek2018quantum} with a symbol-wise entanglement fidelity as distortion measure.
The limitation to iid arises since such symbol-wise entanglement fidelity can be defined only when the quantum state to be encoded is a tensor product of one-qudit states.
The proposed quantum $W_1$ distance does not have this limitation and recovers the Hamming distance for vectors of the canonical basis, and is therefore a candidate to extend quantum rate distortion theory beyond the iid regime.

\item {\bf Quantum differential privacy}

A quantum measurement is gentle if the pre- and post-measurement states are close.
 Ref. \cite{aaronson2019gentle} defines a measurement of the state of $n$ qudits to be differentially private if the probability distributions of the outcome of the measurement performed on any couple of neighboring states are close, \emph{i.e.}, if the measurement cannot distinguish between any two neighboring states.
For product states, the two properties above are intimately connected: any measurement that is gentle on product states is also differentially private and vice versa.
The proposed quantum $W_1$ distance can be thought as a generalization of the notion of neighboring quantum states, and is therefore a candidate to extend beyond product states the connection between quantum differential privacy and gentleness.

  \item {\bf Mixing time of quantum Markov semigroups}

In \autoref{sec:ch}, we have determined upper bounds to the contraction coefficient of the $n$-th tensor power of a one-qudit quantum channel with respect to the proposed quantum $W_1$ distance and we have shown that, in contrast to the situation for the trace distance, such coefficient remains nontrivial in the limit $n\to\infty$.
It is natural to generalize these observations and consider the mixing times of general quantum Markov semigroups with respect to the quantum $W_1$ distance.
A nice property of this approach, in contrast to the bounds derived using the quantum relative entropy, is that the stationary state of the quantum Markov process does not need to have full rank.

\item {\bf Shallow quantum circuits}

The Hamming distance plays a key role in the study of the computational capabilities of quantum circuits \cite{eldar2017local,bravyi2020obstacles}.
The proposed quantum $W_1$ distance recovers the Hamming distance for vectors of the canonical basis and is stable with respect to the action of local shallow quantum circuits.
Therefore, the proposed distance might be useful in characterizing the states generated by constant depth circuits, and it may be able to extend the current results on their computational capabilities.

\item {\bf Quantum many-body Hamiltonians}

In \autoref{prop:lipH}, we have proved that local quantum Hamiltonians have a small quantum Lipschitz constant.
Therefore, the notion of quantum Lipschitz constant can provide a generalization of the notion of local Hamiltonian and lead to the consequent extension of the related properties.
\end{itemize}

\appendices
\section{Alternative definition of neighboring quantum states}\label{app:neigh}
Ref. \cite{aaronson2019gentle} defines the quantum states $\rho,\,\sigma\in\mathcal{S}_n$ to be neighboring if there is a quantum channel $\Phi$ on $\mathcal{H}_n$ that acts on only one qudit and such that either $\rho=\Phi(\sigma)$ or $\sigma=\Phi(\rho)$.
Proceeding along the same lines of \autoref{sec:dist}, this alternative definition of neighboring quantum states induces an alternative quantum $W_1$ norm $\left\|\cdot\right\|_{\tilde{W}_1}$.
In the following \autoref{prop:neigh}, we prove that the norms $\left\|\cdot\right\|_{\tilde{W}_1}$ and $\left\|\cdot\right\|_{W_1}$ are equivalent.

\begin{prop}\label{prop:neigh}
For any $X\in\mathcal{O}_n^T$,
\begin{equation}
\left\|X\right\|_{W_1} \le \left\|X\right\|_{\tilde{W}_1} \le 2\,\frac{d^2-1}{d^2}\left\|X\right\|_{W_1}\,.
\end{equation}
\begin{proof}
Let
\begin{align}
\tilde{\mathcal{N}}_n &= \left\{\pm\left(\rho - \Phi(\rho)\right):\rho\in\mathcal{S}_n,\right.\nonumber\\
& \phantom{=} \quad \left.\Phi\,\textnormal{quantum channel acting on one qudit}\right\}
\end{align}
be the set of the differences between couples of neighboring states according to the alternative definition.
Since $\tilde{\mathcal{N}}_n\subseteq\mathcal{N}_n$, we have
\begin{equation}
\left\|X\right\|_{W_1} \le \left\|X\right\|_{\tilde{W}_1}\,.
\end{equation}
Let $\rho,\,\sigma\in\mathcal{S}_n$ such that $\rho-\sigma\in\mathcal{N}_n$.
Then, there is $i\in[n]$ such that $\mathrm{Tr}_i\rho = \mathrm{Tr}_i\sigma$.
Without loss of generality, we can assume that $i=1$.
We have
\begin{equation}
\left\|\rho-\sigma\right\|_{\tilde{W}_1} \le \left\|\rho - \frac{\mathbb{I}_d}{d}\otimes\mathrm{Tr}_1\rho\right\|_{\tilde{W}_1} + \left\|\frac{\mathbb{I}_d}{d}\otimes\mathrm{Tr}_1\sigma - \sigma\right\|_{\tilde{W}_1}\,.
\end{equation}
Let $U^{(1)},\,\ldots,\,U^{\left(d^2\right)}$ be as in \eqref{eq:Ui}.
Then,
\begin{align}
&\left\|\rho - \frac{\mathbb{I}_d}{d}\otimes\mathrm{Tr}_1\rho\right\|_{\tilde{W}_1}=\nonumber\\
&\frac{1}{d^2}\left\|\sum_{i=2}^{d^2}\left(\rho - \left(U^{(i)}\otimes\mathbb{I}_d^{\otimes\left(n-1\right)}\right)\rho \left({U^{(i)}}^\dag\otimes\mathbb{I}_d^{\otimes\left(n-1\right)}\right)\right)\right\|_{\tilde{W}_1}\nonumber\\
&\le \frac{1}{d^2}\sum_{i=2}^{d^2}\left\|\rho - \left(U^{(i)}\otimes\mathbb{I}_d^{\otimes\left(n-1\right)}\right)\rho \left({U^{(i)}}^\dag\otimes\mathbb{I}_d^{\otimes\left(n-1\right)}\right)\right\|_{\tilde{W}_1}\nonumber\\
&\le \frac{d^2-1}{d^2}\,,
\end{align}
therefore
\begin{equation}
\left\|\rho-\sigma\right\|_{\tilde{W}_1} \le 2\,\frac{d^2-1}{d^2}\,.
\end{equation}
Then,
\begin{equation}
\left\|X\right\|_{\tilde{W}_1} \le 2\,\frac{d^2-1}{d^2}\left\|X\right\|_{W_1}\,,
\end{equation}
and the claim follows.
\end{proof}
\end{prop}

\section{Efficient estimation of the quantum Lipschitz constant}\label{app:a}
The following \autoref{prop:L} provides an estimate of the quantum Lipschitz constant up to a multiplicative error $\sqrt{2}$ that does not require any optimization.
\begin{prop}[Efficient estimation of the quantum Lipschitz constant]\label{prop:L}
For any $i\in[n]$, let $\mathcal{E}_i$ be the quantum channel on $\mathcal{H}_n$ that replaces the state of the $i$-th qudit with the maximally mixed state.
Then, for any $H\in\mathcal{O}_n$,
\begin{align}
&\frac{d^2}{d^2-1}\max_{i\in[n]}\left\|H - \mathcal{E}_i(H)\right\|_\infty \le \left\|H\right\|_L\,,\nonumber\\
&\left\|H\right\|_L \le 2\max_{i\in[n]}\left\|H - \mathcal{E}_i(H)\right\|_\infty\,.
\end{align}
\begin{proof}
Let $X\in\mathcal{N}_n$.
Then, there exists $i\in[n]$ such that $\mathrm{Tr}_iX = 0$, and
\begin{equation}
0 = \mathrm{Tr}\left[H\,\mathcal{E}_i(X)\right] = \mathrm{Tr}\left[\mathcal{E}_i(H)\,X\right]\,.
\end{equation}
We then have
\begin{align}
\mathrm{Tr}\left[H\,X\right] &= \mathrm{Tr}\left[\left(H - \mathcal{E}_i(H)\right)X\right] \le \left\|H - \mathcal{E}_i(H)\right\|_\infty\left\|X\right\|_1\nonumber\\
&\le 2\left\|H - \mathcal{E}_i(H)\right\|_\infty\,,
\end{align}
therefore
\begin{equation}
\left\|H\right\|_L \le 2\max_{i\in[n]}\left\|H - \mathcal{E}_i(H)\right\|_\infty\,.
\end{equation}

Let $i\in[n]$, and let $\rho\in\mathcal{S}_n$ such that
\begin{align}
\left|\mathrm{Tr}\left[H\left(\rho - \mathcal{E}_i(\rho)\right)\right]\right| &= \left|\mathrm{Tr}\left[\left(H - \mathcal{E}_i(H)\right)\rho\right]\right|\nonumber\\
&= \left\|H - \mathcal{E}_i(H)\right\|_\infty\,.
\end{align}
From \eqref{eq:1Ti} and \autoref{lem:E} of \autoref{app:lemmas},
\begin{equation}
\left\|\rho - \mathcal{E}_i(\rho)\right\|_{W_1} = \frac{1}{2}\left\|\rho - \mathcal{E}_i(\rho)\right\|_1 \le \frac{d^2-1}{d^2}\,,
\end{equation}
therefore
\begin{equation}
\left\|H\right\|_L \ge \frac{d^2}{d^2-1}\left\|H - \mathcal{E}_i(H)\right\|_\infty\,,
\end{equation}
and the claim follows.
\end{proof}
\end{prop}

\section{Lemmas}\label{app:lemmas}

\begin{lem}\label{lem:equiv}
For any $X\in\mathcal{O}_n^T$,
\begin{align}\label{eq:W1lem}
\left\|X\right\|_{W_1} &= \frac{1}{2}\min\left(\sum_{i=1}^n\left\|X^{(i)}\right\|_1:X^{(i)}\in\mathcal{O}_n^T,\,\mathrm{Tr}_i\,X^{(i)}=0,\right.\nonumber\\
&\phantom{= \frac{1}{2}\min} \quad \left.X=\sum_{i=1}^nX^{(i)}\right)\,.
\end{align}
\begin{proof}
Throughout this proof, $\left\|\cdot\right\|_{W_1}$ denotes the norm defined in \eqref{eq:W1lem}.
The optimization in \eqref{eq:W1lem} is performed over a compact set, therefore the minimum is achieved.
To prove the claim, it is sufficient to prove that the unit ball of $\left\|\cdot\right\|_{W_1}$ coincides with $\mathcal{B}_n$.

On the one hand, let $X\in\mathcal{B}_n$.
Since each $\mathcal{N}_n^{(i)}$ is convex, $X$ is a convex combination of $n$ elements, each belonging to the corresponding $\mathcal{N}_n^{(i)}$, \emph{i.e.}, there exists a probability distribution $p$ on $[n]$ such that
\begin{align}
&X = \sum_{i=1}^n p_i\left(\rho^{(i)} - \sigma^{(i)}\right)\,,\nonumber\\
&\rho^{(i)},\,\sigma^{(i)}\in\mathcal{S}_n\,,\qquad \mathrm{Tr}_i\rho^{(i)} = \mathrm{Tr}_i\sigma^{(i)}\,.
\end{align}
Therefore, choosing in \eqref{eq:W1lem}
\begin{equation}
X^{(i)} = p_i\left(\rho^{(i)} - \sigma^{(i)}\right)\,,
\end{equation}
we get
\begin{equation}
\left\|X\right\|_{W_1} \le \frac{1}{2}\sum_{i=1}^np_i\left\|\rho^{(i)} - \sigma^{(i)}\right\|_1 \le 1\,,
\end{equation}
and $X$ belongs to the unit ball of $\left\|\cdot\right\|_{W_1}$.

On the other hand, let $X\in\mathcal{O}_n^T$ such that $\left\|X\right\|_{W_1}=1$.
Then, there exist $X^{(1)},\,\ldots X^{(n)}$ as in \eqref{eq:W1lem} such that
\begin{equation}
\frac{1}{2}\sum_{i=1}^n\left\|X^{(i)}\right\|_1 = 1\,.
\end{equation}
For any $i\in[n]$, let
\begin{equation}
p_i = \frac{1}{2}\left\|X^{(i)}\right\|_1\,,
\end{equation}
such that $p$ is a probability distribution on $[n]$.
We can express each $X^{(i)}$ as
\begin{equation}
X^{(i)} = p_i\left(\rho^{(i)} - \sigma^{(i)}\right)\,,
\end{equation}
where $\rho^{(i)},\,\sigma^{(i)}\in\mathcal{S}_n$ have orthogonal supports.
Since $\mathrm{Tr}_iX^{(i)}=0$, $\rho^{(i)}$ and $\sigma^{(i)}$ are neighboring, and $\rho^{(i)} - \sigma^{(i)}\in\mathcal{N}_n$.
Since
\begin{equation}
X = \sum_{i=1}^n X^{(i)} = \sum_{i=1}^n p_i\left(\rho^{(i)} - \sigma^{(i)}\right)\,,
\end{equation}
$X\in\mathcal{B}_n$, and the claim follows.
\end{proof}
\end{lem}

\begin{lem}\label{lem:XOY}
For any $X\in\mathcal{O}_m^T$ and any $Y\in\mathcal{O}_n$,
\begin{equation}
\left\|X\otimes Y\right\|_{W_1} \le \left\|X\right\|_{W_1}\left\|Y\right\|_1\,.
\end{equation}
\begin{proof}
Let $X^{(1)},\,\ldots,\,X^{(m)}\in\mathcal{O}_m^T$ such that
\begin{equation}
\mathrm{Tr}_iX^{(i)}=0\quad\forall\,i\in[m]\,,\qquad X = \sum_{i=1}^mX^{(i)}\,.
\end{equation}
Since
\begin{equation}
X\otimes Y = \sum_{i=1}^mX^{(i)}\otimes Y\,,
\end{equation}
we get
\begin{equation}
\left\|X\otimes Y\right\|_{W_1} \le \frac{1}{2}\sum_{i=1}^m\left\|X^{(i)}\right\|_1\left\|Y\right\|_1\,,
\end{equation}
and the claim follows.
\end{proof}
\end{lem}

\begin{lem}\label{lem:E}
For any $X\in\mathcal{O}_n$,
\begin{equation}
\left\|X - \frac{\mathbb{I}_d}{d}\otimes\mathrm{Tr}_1X\right\|_1 \le 2\,\frac{d^2-1}{d^2}\left\|X\right\|_1\,.
\end{equation}
\begin{proof}
Let $U^{(1)},\,\ldots,\,U^{\left(d^2\right)}$ be a set of unitary operators on $\mathbb{C}^d$ such that
\begin{equation}\label{eq:Ui}
U^{(1)} = \mathbb{I}_d\,,\qquad \mathrm{Tr}\left[{U^{(i)}}^\dag U^{(j)}\right] = d\,\delta_{ij}\,,\qquad i,\,j\in\left[d^2\right]\,.
\end{equation}
Then,
\begin{align}
&\left\|X - \frac{\mathbb{I}_d}{d}\otimes\mathrm{Tr}_1X\right\|_1\nonumber\\
&= \frac{1}{d^2}\left\|\left(d^2-1\right)X\phantom{\sum_{i=2}^{d^2}}\right.\nonumber\\
&\phantom{= \frac{1}{d^2}} \quad \left. - \sum_{i=2}^{d^2}\left(U^{(i)}\otimes\mathbb{I}_d^{\otimes\left(n-1\right)}\right)X \left({U^{(i)}}^\dag\otimes\mathbb{I}_d^{\otimes\left(n-1\right)}\right)\right\|_1\nonumber\\
&\le 2\,\frac{d^2-1}{d^2}\left\|X\right\|_1\,,
\end{align}
and the claim follows.
\end{proof}
\end{lem}

\begin{lem}\label{lem:EPR}
Let $\gamma$ be a maximally entangled state of $\left(\mathbb{C}^d\right)^{\otimes2}$.
Then, for any even $n$,
\begin{equation}
\left\|\gamma^{\otimes\frac{n}{2}} - \frac{\mathbb{I}_d^{\otimes n}}{d^n}\right\|_{W_1} = \frac{n}{2}\,\frac{d^2-1}{d^2}\,.
\end{equation}
\begin{proof}
From \autoref{prop:tens}, it is sufficient to prove the claim for $n=2$.
Since
\begin{equation}
\mathrm{Tr}_1\gamma = \frac{\mathbb{I}_d}{d}\,,
\end{equation}
we have from \eqref{eq:1Ti}
\begin{equation}
\left\|\gamma - \frac{\mathbb{I}_d^{\otimes 2}}{d^2}\right\|_{W_1} = \frac{1}{2}\left\|\gamma - \frac{\mathbb{I}_d^{\otimes 2}}{d^2}\right\|_{1} = \frac{d^2-1}{d^2}\,,
\end{equation}
and the claim follows.
\end{proof}
\end{lem}

\begin{lem}\label{lem:S}
Let $\rho$ and $\sigma$ be quantum states of the Hilbert space $\mathcal{H}_A\otimes\mathcal{H}_B$ such that $\rho_B = \sigma_B$.
Then,
\begin{equation}
\left|S(\rho) - S(\sigma)\right| \le 2\ln\dim\mathcal{H}_A\,.
\end{equation}
\begin{proof}
We have
\begin{align}
\left|S(\rho) - S(\sigma)\right| &= \left|S(\rho_B) + S(A|B)_{\rho} - S(\sigma_B) - S(A|B)_{\sigma}\right|\nonumber\\
&= \left|S(A|B)_{\rho} - S(A|B)_{\sigma}\right|\nonumber\\
&\le \left|S(A|B)_{\rho}\right| + \left|S(A|B)_{\sigma}\right| \le 2\ln\dim\mathcal{H}_A\,,
\end{align}
and the claim follows.
\end{proof}
\end{lem}

\begin{lem}\label{lem:nc}
Let $p$ and $q$ be probability distributions on $[d]^n$ whose marginals over the first $n-1$ components coincide, \emph{i.e.}, such that $p(x_1\ldots x_{n-1}) = q(x_1\ldots x_{n-1})$ for any $x_1,\,\ldots,\,x_{n-1}\in[d]$.
Then,
\begin{equation}
W_1(p,q) \le 1\,.
\end{equation}
\begin{proof}
We have for any $x\in[d]^n$
\begin{align}
p(x) &= p(x_1\ldots x_{n-1})\,p(x_n|x_1\ldots x_{n-1})\,,\nonumber\\
q(x) &= q(x_1\ldots x_{n-1})\,q(x_n|x_1\ldots x_{n-1})\nonumber\\
&= p(x_1\ldots x_{n-1})\,q(x_n|x_1\ldots x_{n-1})\,.
\end{align}
Since the $W_1$ distance is jointly convex, we have
\begin{align}
W_1(p,q) &\le \sum_{x_1,\,\ldots,\,x_{n-1}\in[d]}p(x_1\ldots x_{n-1})\,\cdot\nonumber\\
&\phantom{\le}\quad \cdot W_1\left(p(\cdot|x_1\ldots x_{n-1}),\,q(\cdot|x_1\ldots x_{n-1})\right)\nonumber\\
&\le 1\,,
\end{align}
and the claim follows.
\end{proof}
\end{lem}

\section{Proof of the \texorpdfstring{$W_1$}{W1} continuity of the Shannon entropy}\label{app:entc}
Let $X,\,Y$ be random variables with values in $[d]^n$ whose joint probability distribution is the optimal coupling between $p$ and $q$.
For any $i\in[n]$, let $p_i$ be the probability that $X_i\neq Y_i$, such that
\begin{equation}
W_1(p,q) = \sum_{i=1}^n p_i\,.
\end{equation}
We have
\begin{align}
S(X) - S(Y) &\overset{\textnormal{(a)}}{\le} S(XY) - S(Y) = S(X|Y)\nonumber\\
&\overset{\textnormal{(b)}}{\le} \sum_{i=1}^nS(X_i|Y) \overset{\textnormal{(c)}}{\le} \sum_{i=1}^nS(X_i|Y_i)\nonumber\\
&\overset{\textnormal{(d)}}{\le} \sum_{i=1}^n\left(h_2(p_i) + p_i\ln\left(d-1\right)\right)\nonumber\\
&\overset{\textnormal{(e)}}{\le} n\,h_2\left(\frac{1}{n}\sum_{i=1}^n p_i\right) + \ln\left(d-1\right)\sum_{i=1}^np_i\nonumber\\
&= n\,h_2\left(\frac{W_1(p,q)}{n}\right) + W_1(p,q)\ln\left(d-1\right)\,,
\end{align}
where (a) follows from the monotonicity of the Shannon entropy, (b) and (c) follow from the strong subadditivity of the Shannon entropy, (d) follows from Fano's inequality and (e) follows from Jensen's inequality applied to the concave function $h_2$.
The claim follows.

\bibliographystyle{IEEEtran}

\bibliography{distance}

\begin{IEEEbiographynophoto}{Giacomo De Palma}
was born in Lanciano (CH), Italy, in 1990.
He received the B.S. and M.S. degrees in physics from the University of Pisa in 2011 and 2013, respectively, and the ``Diploma di Licenza'' and Ph.D. degrees in physics from Scuola Normale Superiore in 2014 and 2016, respectively.

He was a Postdoc and a Marie-Sk\l odowska Curie Fellow at the University of Copenhagen from 2016 to 2018 and from 2018 to 2019, respectively.
From 2019 to 2021, he was a Postdoctoral Associate at MIT.
Since 2021, he has been a Tenure-track Assistant Professor in mathematical physics at Scuola Normale Superiore.
He is the author of more than 30 scientific articles.
His research interests include all aspects of quantum information and quantum machine learning.

Dr. De Palma is a member of the International Association of Mathematical Physics (IAMP), and was a recipient of the Best Italian Researcher in Denmark (BIRD) award in 2018.
\end{IEEEbiographynophoto}

\begin{IEEEbiographynophoto}{Milad Marvian}
completed his Ph.D. in 2017 at the University of Southern California.

He was a Postdoctoral Associate at MIT from 2018 to 2020. Since 2020, he is an Assistant Professor in the Department of Electrical \& Computer Engineering at the University of New Mexico and also a member of the Center for Quantum Information and Control (CQuIC).
His research interest includes quantum algorithms and quantum machine learning, quantum error correction, and open quantum systems.
\end{IEEEbiographynophoto}

\begin{IEEEbiographynophoto}{Dario Trevisan}
was born in Mirano (VE), Italy, on June 15, 1987. He received both the B.S.~and M.S.~degrees in mathematics from the University of Pisa, in 2009 and 2011, respectively, and the Ph.D.~degree in mathematics from Scuola Normale Superiore in 2014.

He is currently Assistant Professor at the University of Pisa.
\end{IEEEbiographynophoto}

\begin{IEEEbiographynophoto}{Seth Lloyd}
was born in Boston, MA, USA in 1960.   He received the B.A. degree in physics from Harvard University in 1982, the master’s degree in mathematics from the University of Cambridge (Part III) in 1983, the M.Phil. degree in history and philosophy of science from the University of Cambridge in 1984, and the Ph.D. degree in physics from The Rockefeller University in 1988.

From 1988 to 1991 he was a Postdoc with Murray Gell-Mann at Caltech, and from 1991 to 1994 a Postdoc at Los Alamos.   Since 1994 he has been Professor of Mechanical Engineering at MIT.  Dr. Lloyd's work focuses on the role of information in the universe, including quantum information and complexity.  He is the author of more than 250 scientific papers and of a book, Programming the Universe.

He is the recipient of the Edgerton Prize and the International Quantum Computation and Communication prize.
\end{IEEEbiographynophoto}

\end{document}